\documentclass[a4paper,10pt]{article}
\usepackage{jheppub}

\usepackage[utf8]{inputenc}
\usepackage{amsmath,amssymb,bm,slashed,braket}
\usepackage{dsfont}
\usepackage{amsfonts}
\usepackage{bm,bbm}
\usepackage{graphicx}
\usepackage{slashed} 
\usepackage[dvipsnames,table]{xcolor}
\usepackage[normalem]{ulem}
\usepackage{soul}
\usepackage{siunitx} 
\usepackage{hyperref}
\hypersetup{colorlinks,citecolor= blue,linkcolor= blue, urlcolor=blue}
\usepackage{ulem}
\usepackage{array}
\usepackage{verbatim}
\usepackage{epsfig}
\usepackage{multirow, booktabs}
\usepackage{bbold}
\usepackage[version=4]{mhchem}
\usepackage[capitalise]{cleveref}
\usepackage{makecell}
\usepackage{orcidlink}

\allowdisplaybreaks[4]
\def\lsim{\mathrel{\raise.3ex\hbox{$<$\kern-.75em\lower1ex\hbox{$\sim$}}}}
\def\gsim{\mathrel{\raise.3ex\hbox{$>$\kern-.75em\lower1ex\hbox{$\sim$}}}}

\newcommand{\calO}{ {\cal O} }

\title{Charged lepton flavor violating decays with a pair of light dark matter and muonium invisible decay}

\author[a,b]{Sahabub Jahedi\,\orcidlink{0000-0003-1016-8264},}
\emailAdd{sahabub@m.scnu.edu.cn}
\affiliation[a]{State Key Laboratory of Nuclear Physics and
Technology, Institute of Quantum Matter, South China Normal
University, Guangzhou 510006, China}
\affiliation[b]{Guangdong Basic Research Center of Excellence for
Structure and Fundamental Interactions of Matter, Guangdong
Provincial Key Laboratory of Nuclear Science, Guangzhou
510006, China}
\author[a,b]{Yi Liao\,\orcidlink{0000-0002-1009-5483},}
\emailAdd{liaoy@m.scnu.edu.cn}
\author[a,b]{Xiao-Dong Ma\,\orcidlink{0000-0001-7207-7793}}
\emailAdd{maxid@scnu.edu.cn}

\abstract{
In this paper, we initiate the study of lepton flavor violating (LFV) dark matter (DM) interactions, 
expanding our focus beyond the flavor-conserving DM interactions typically considered in conventional direct and indirect detections.  
We work in an effective field theory (EFT) framework, focusing on the leading-order local operators of the form, $\bar \ell_j \Gamma \ell_i\,{\tt DM}^2$, where $(ij)=(e\mu, e\tau, \mu\tau)$ and the DM includes the three well-known scenarios: a scalar, a fermion, and a vector.
We derive the invariant-mass distribution for the three-body decay $\ell_i \to \ell_j +{\tt DM+DM}$ and demonstrate that it can be used to distinguish between different operator structures and to determine the DM mass.
By utilizing current experimental bounds on the charged muon LFV decay involving neutrinos and the ratio of tau leptonic decay widths, we establish stringent limits on the effective scale associated with each operator. 
Additionally, for the $e\mu$ flavor combination, we investigate the muon four-body radiative decay ($\mu\to e +{\tt DM+DM}+\gamma$) to complement our probe of such interactions. 
Finally, we examine muonium invisible decays based on the derived bounds on the effective operators and find that the branching ratios can be significantly enhanced compared to the predictions of the standard model.
In particular, any future observation of the para-muonium invisible decay serves as a compelling signature for these flavored DM interactions. 

}

\makeatletter
\gdef\@fpheader{}
\makeatother
\begin{document} 

\maketitle
\setcounter{page}{2}
\section{Introduction}
The non-zero neutrino masses and unknown nature of dark matter (DM) provide two compelling indications of physics beyond the Standard Model (SM).
Among numerous proposed beyond-SM scenarios, an especially appealing approach is to address these two seemingly disparate issues within a unified framework. 
Many new physics (NP) models have been developed based on this idea, including the well-known scotogenic neutrino mass model \cite{Tao:1996vb,Ma:2006km} as well as other similar frameworks.
Generally, such models give rise to DM-lepton interactions, including both the lepton flavor conserving (LFC) and lepton flavor violating (LFV) interactions between SM particles and the DM. 
While the LFC interactions between a pair of DM fields and a pair of electron (muon or tau) fields have been extensively studied in recent years from both DM direct and indirect detection experiments (see the recent review papers \cite{Schumann:2019eaa,Lin:2019uvt,PerezdelosHeros:2020qyt,Billard:2021uyg,Cirelli:2024ssz} and references therein), the LFV interactions involving SM particles and the DM have not received comparative attention. 
Investigating these LFV couplings not only deepens our understanding of the relationship between DM and neutrinos but also opens up new avenues for experimental investigation. 

In this work, we will initiate the study of LFV interactions related to the DM particle, focusing specifically on the sub-GeV scale DM. 
The motivation for considering such a light DM candidate is primarily driven by the challenges associated with the DM direct detection through nuclear recoil, as there are strong constraints on DM candidates with masses greater than tens of GeV \cite{LZ:2024zvo}. 
For the LFV interactions involving a pair of light DM, 
the charged LFV (cLFV) decays with a DM pair in the final state serve as the most effective observables to probe these flavored DM interactions. 

In the quark sector, various studies have explored flavor-changing neutral current (FCNC) decays of mesons (baryons) into a light meson (baryon) plus a pair of DM or dark sector particles \cite{Kamenik:2011vy,Altmannshofer:2009ma,Li:2020dpc,Bird:2004ts,Tandean:2019tkm,Li:2019cbk,He:2021yoz,He:2022ljo,He:2023bnk,He:2024iju,Buras:2024ewl}. However, corresponding studies in the lepton sector have received limited attention, both theoretically and experimentally.  Only a few papers have concentrated on the cLFV two-body decays involving a single invisible particle, such as a light (pseudo)scalar $\ell_i \to \ell_j + a(+\gamma)$ \cite{Jodidio:1986mz,Paradisi:2005tk,Bjorkeroth:2018dzu,Calibbi:2020jvd,Cheung:2021mol,Bauer:2021mvw,Jho:2022snj,Banerjee:2022xuw,Calibbi:2024rcm,Ema:2025bww,Jiang:2025nie,Knapen:2024fvh,Greljo:2025ljr,Bigaran:2025uzn} or a dark photon $\ell_i \to \ell_j + X$ \cite{Zhevlakov:2023jzt}, within the framework of the effective field theory (EFT) or simplified models.  
In this work, we aim to bridge this gap by systematically investigating the three cLFV decays that involve a pair of DM particles ($\mu \to e +{\tt DM +DM}$, $\tau \to e +{\tt DM +DM}$, and $\tau \to \mu +{\tt DM +DM}$) within a general EFT framework.\footnote{
The cLFV decays involving a neutrino pair ($\ell_i \to \ell_j + \nu_\alpha \nu_\beta$) represent another similar possibility that can be systematically analyzed within the framework of the SMEFT \cite{Grzadkowski:2010es} or the $\nu$SMEFT \cite{Liao:2016qyd}. } 
Focusing on a pair of DM particles rather than a single one is mainly motivated by DM stability, which can be naturally guaranteed by an underlying symmetry that prevents DM  decay and gives rise to the interactions under consideration.
Additionally, for the $e\mu$ flavor combination, we include the four-body radiative decay process $\mu \to e +\gamma+{\tt DM +DM}$ due to its experimental relevance.

Experimentally, the cLFV decays with visible particles in the final state, such as $\mu\to e\gamma$ and $\mu\to 3 e$, have been the focus of extensive searches over many years \cite{MEG:2013mmu,SINDRUM:1987nra}, resulting in very stringent bounds on the corresponding branching ratios. 
Currently, there are many proposed future experiments (including the MEG II \cite{MEGII:2018kmf}, 
Mu2e \cite{Mu2e:2014fns}, 
COMET \cite{COMET:2018auw}, and 
STCF \cite{Achasov:2023gey}) that are dedicated to investigate these processes with high precision. Search for an invisible light boson has been conducted through the cLFV tau decays at both the Belle \cite{Belle:2025bpu} and Belle II \cite{Belle-II:2022heu} experiments. 
Since the processes we consider, $\ell_i \to \ell_j +{\tt DM +DM}$ ($\mu \to e+{\tt DM +DM} +\gamma $), are analogous to the SM processes $\ell_\alpha^- \to \ell_\beta^-  \nu_\alpha  \bar\nu_\beta$ ($\mu^- \to e^- \nu_\mu \bar \nu_e \gamma$), it is anticipated that the aforementioned experiments can also be used to search for these processes. 
In the absence of direct constraints on them and considering the similarities between DM and neutrinos in the experimental detection, we will utilize the current most stringent experimental upper limits on the branching ratios of analogous LFV decays of $\mu$ and $\tau$ involving neutrinos or dark sector particles to set bounds on the parameters in the EFT framework. Numerically, we take the PDG number (90\% C.L.) for the LFV muon decay process,
$\mathcal{B}(\mu^- \to e^- \nu_e \bar\nu_{\mu}) < 0.012$ \cite{ParticleDataGroup:2024cfk}.   
Regarding tau decays, we utilize the ratio of the leptonic branching fractions $R^\tau_{\mu/e}=\mathcal{B}(\tau \to \mu + {\rm inv.})/\mathcal{B}(\tau \to e + {\rm inv.})$ to derive the corresponding limits. In the SM, this ratio is precisely predicted as $R^{\tau,\tt SM}_{\mu/e}=0.972563(3)$~\cite{Pich:2024qob}. Currently, the  world average value is $R^{\tau,\tt ave.}_{\mu/e}=0.9730(22)$, which combines the PDG value 0.9762(28)~\cite{ParticleDataGroup:2024cfk} and the recent Belle II result 0.9675(37)~\cite{Belle-II:2024vvr}.
We define the NP contribution as $R^{\tau,\tt NP}_{i/e}=\mathcal{B}(\tau \to i +{\tt DM +DM})/\mathcal{B}(\tau^- \to e^-\nu_\tau\bar\nu_e)_{\tt SM}$ with $i=e,~\mu$ so that
$R^\tau_{\mu/e}\approx R^{\tau,\tt SM}_{\mu/e} +R^{\tau,\tt NP}_{\mu/e} -R^{\tau,\tt NP}_{e/e} $.
In our analysis, we quantify the room for new physics in these channels as the 90\% confidence level limit derived from the difference between the experimental average and the SM expectation.\footnote{
It should be noted that the limits derived here are approximate and subject to two limitations. First, variations in operator structures may lead to differences in kinematic distributions and experimental efficiencies compared to the SM $V-A$ current structure, thus potentially affecting the resulting bounds. 
Second, we assume that only one DM channel---either $i=e$ or $i=\mu$---is open a time, and we neglect the possibility that when both channels are open, cancellations in $R^\tau_{\mu/e}$ might occur to maintain consistency with the SM.}
This yields the bounds: $\mathcal{B}(\tau \to e+{\tt DM +DM}) < 5.8 \times 10^{-4} $ and $\mathcal{B}(\tau \to \mu+{\tt DM +DM}) < 7.2 \times 10^{-4} $,
where we have used the SM prediction $\mathcal{B}(\tau^- \to e^- \nu_e \bar\nu_{\tau})_{\tt SM}=17.785\%$~\cite{Pich:2024qob}. 

For the four-body radiative mode, we take the experimental uncertainties in the branching ratio of $\mu^- \to e^- \bar{\nu}_e \nu_\mu \gamma $ as the NP window, for which
$\mathcal{B}(\mu^- \to e^-\bar \nu_e  \nu_\mu \gamma ) = 6.0(5) \times 10^{-8}$~\cite{ParticleDataGroup:2024cfk},
with experimental cuts of $E_e>45\,\rm MeV$ and $E_\gamma > 40\,\rm MeV$.
In addition to the aforementioned free lepton decays, the bound muonium atom ($M_\mu$) can also be employed to explore these interactions. Currently, the proposed MACE experiment \cite{Bai:2024skk} aims to investigate the muonium-antimuonium oscillation, along with other new physics phenomena, including muonium invisible decays $M_\mu \to \rm inv.$. Therefore, we also consider the muonium invisible two-body decay process ($M_\mu \to {\tt DM+DM}$) in this study.  

EFT is a simple yet powerful approach for investigating NP effects beyond the SM predictions, due to its model-agnostic nature. For a light DM particle, its interactions with the SM particles at an energy scale far below the electroweak scale can be well described by the low energy EFT extended by a light DM particle, referred to as the dark sector EFT (DSEFT) \cite{He:2022ljo,Liang:2023yta}. 
In this work, we will operate within this DSEFT framework and consider three classical DM scenarios with spin up to one: a scalar $\phi$, a fermion $\chi$, and a vector $X$. We will first collect all the leading-order local effective operators ($ (\bar\ell_i \Gamma \ell_j){\tt DM}^2$) for the three DM cases. In terms of the DSEFT operators, we then study the three-body processes $\ell_i \to \ell_j +{\tt DM +DM}$, including the invariant-mass ($q^2$) distribution of the DM pair and the current experimental bound on the effective scale associated with each operator. 
The $q^2$-distribution serves as a valuable observable that can help distinguish the operator structures in future experiments. For the $e\mu$ flavor combination, we also examine the four-body radiative decay process $\mu\to e +{\tt DM+DM} +\gamma$ by analyzing the missing energy distribution and the constraints on EFT interactions from current experimental data.  
Given the interest in the muonium atom, we additionally evaluate the sensitivity of the branching ratios for muonium invisible decays by using constraints derived from muon decays. 
Our study indicates that the effective scale can be probed up to the TeV scale for dimension-6 (dim-6) operators. 
Notably, the bounds on muonium two-body invisible decays arising from the insertion of several operators can surpass that of the SM process with a neutrino pair, presenting opportunities for further investigation in future experiments.

The paper is structured as follows.
In \cref{sec:DSEFT}, we describe the EFT framework and collect the relevant effective operators responsible for the processes under consideration. \cref{sec:3Bdecay} is dedicated to analyzing all possible three-body decays ($\ell_i \to \ell_j +{\tt DM}+{\tt DM}$), including the invariant mass distribution and the constraints on the effective scale derived from current lepton decay data. In \cref{sec:4Bdecay}, we further investigate the muon four-body radiative decay and its constraints, complementing those obtained from the three-body decay mode. Based on the constraints on the effective scale established in the previous sections, we explore the implications for the invisible decay modes of the muonium atom in \cref{sec:Mmudecay}. Our summary is presented in \cref{sec:summary}. 
Additionally, \cref{app:Mi2jDMDM} summarizes the matrix elements for the three-body decays and \cref{app:diff.rate} collects the $q^2$-distribution formulas.

\section{Lepton-DM interactions in the DSEFT}
\label{sec:DSEFT}

\begin{table}
\center
\resizebox{\linewidth}{!}{
\renewcommand\arraystretch{1.}
\begin{tabular}{| c | l | c | l |}
\hline
\multicolumn{4}{|c|}{\bf Scalar DM case}
\\
\hline
$\calO_{\ell\phi}^{{\tt S},ji}$  &  $ (\bar{\ell}_j\ell_i )(\phi^\dagger \phi)$ & 
$\calO_{\ell\phi}^{{\tt P},ji}$ &  $  (\bar{\ell}_j i \gamma_5\ell_i )(\phi^\dagger \phi)$
\\
\hline
$\calO_{\ell\phi}^{{\tt V},ji}$ & 
$(\bar{\ell}_j\gamma^\mu\ell_i ) 
(\phi^\dagger i \overleftrightarrow{\partial_\mu} \phi) \, (\times) $ & 
$ \calO_{\ell\phi}^{{\tt A},ji}$  & 
$(\bar{\ell}_j\gamma^\mu\gamma_5\ell_i ) 
(\phi^\dagger i \overleftrightarrow{\partial_\mu} \phi)  \, (\times)$  
\\
\hline
\multicolumn{4}{|c|}{\bf Fermion DM case}
\\
\hline
$\calO_{\ell\chi1}^{{\tt S},ji} $ &
$ (\bar{\ell}_j\ell_i )(\overline{\chi}\chi)$ &
$\calO_{\ell\chi1}^{{\tt P},ji} $ &
$  (\bar{\ell}_j i \gamma_5\ell_i )(\overline{\chi}\chi)$
\\\hline
$\calO_{\ell\chi2}^{{\tt S},ji} $ & 
$(\bar{\ell}_j\ell_i )(\overline{\chi}i \gamma_5\chi)$ &
$\calO_{\ell\chi2}^{{\tt P},ji} $ & 
$ (\bar{\ell}_j \gamma_5\ell_i )(\overline{\chi} \gamma_5\chi)$
\\\hline
$\calO_{\ell\chi1}^{{\tt V},ji} $ &
$  (\bar{\ell}_j\gamma^\mu \ell_i )(\overline{\chi}\gamma_\mu  \chi)\, (\times)$ &
$\calO_{\ell\chi1}^{{\tt A},ji} $ &
$(\bar{\ell}_j\gamma^\mu\gamma_5 \ell_i )(\overline{\chi}\gamma_\mu \chi)\, (\times)$
\\\hline
$\calO_{\ell\chi2}^{{\tt V},ji} $ &
$ (\bar{\ell}_j\gamma^\mu\ell_i )(\overline{\chi}\gamma_\mu  \gamma_5\chi)$  &
$\calO_{\ell\chi2}^{{\tt A},ji} $ &
$(\bar{\ell}_j\gamma^\mu\gamma_5\ell_i )(\overline{\chi}\gamma_\mu \gamma_5\chi)$
\\\hline
$\calO_{\ell\chi1}^{{\tt T},ji} $ &
$ (\bar{\ell}_j\sigma^{\mu\nu} \ell_i )(\overline{\chi}\sigma_{\mu\nu}\chi)\,(\times)$ &
$\calO_{\ell\chi2}^{{\tt T},ji} $ &
$ (\bar{\ell}_j\sigma^{\mu\nu}\ell_i )(\overline{\chi}\sigma_{\mu\nu} \gamma_5\chi)\,(\times)$
\\\hline 
\multicolumn{4}{|c|}{\bf Vector DM case A}
\\
\hline
$\calO_{\ell X}^{{\tt S},ji} $ &
$ (\bar{\ell}_j\ell_i )(X_\mu^\dagger X^\mu)$ &
$ \calO_{\ell X}^{{\tt P},ji} $ &
$ (\bar{\ell}_ji \gamma_5\ell_i )(X_\mu^\dagger X^\mu)$
\\\hline
$ \calO_{\ell X1}^{{\tt T},ji} $ & 
$ {i \over 2} (\bar{\ell}_j \sigma^{\mu\nu}\ell_i) (X_\mu^\dagger X_\nu - X_\nu^\dagger X_\mu)  \, (\times) $ &
$ \calO_{\ell X2}^{{\tt T},ji} $ &
$ {1\over 2} (\bar{\ell}_j\sigma^{\mu\nu}\gamma_5\ell_i)(X_\mu^\dagger X_\nu - X_\nu^\dagger X_\mu)\,(\times) $
\\\hline
$ \calO_{\ell X1}^{{\tt V},ji} $ &
$ {1\over 2} [ \bar{\ell}_j\gamma_{(\mu} i \overleftrightarrow{D_{\nu)} } \ell_i ] (X^{\mu \dagger} X^\nu + X^{\nu \dagger} X^\mu)$ & 
$\calO_{\ell X1}^{{\tt A},ji} $ & 
${1\over 2} [\bar{\ell}_j\gamma_{(\mu} \gamma_5 i \overleftrightarrow{D_{\nu)} }  \ell ](X^{\mu \dagger} X^\nu + X^{\nu \dagger} X^\mu  )$
\\\hline 
$ \calO_{\ell X2}^{{\tt V},ji} $ &
$ (\bar{\ell}_j\gamma_\mu\ell_i)\partial_\nu (X^{\mu \dagger} X^\nu + X^{\nu \dagger} X^\mu)$ & 
$\calO_{\ell X2}^{{\tt A},ji} $ &
$(\bar{\ell}_j\gamma_\mu \gamma_5\ell_i )\partial_\nu (X^{\mu \dagger} X^\nu + X^{\nu \dagger} X^\mu)$
\\\hline
$ \calO_{\ell X3}^{{\tt V},ji} $ &
$ (\bar{\ell}_j\gamma_\mu\ell_i )( X_\rho^\dagger \overleftrightarrow{\partial_\nu} X_\sigma )\epsilon^{\mu\nu\rho\sigma}$ & 
$\calO_{\ell X3}^{{\tt A},ji} $ &
$(\bar{\ell}_j\gamma_\mu\gamma_5\ell_i ) (X_\rho^\dagger \overleftrightarrow{ \partial_\nu} X_\sigma )\epsilon^{\mu\nu\rho\sigma}$
\\\hline
$ \calO_{\ell X4}^{{\tt V},ji} $ & 
$ (\bar{\ell}_j\gamma^\mu\ell_i )(X_\nu^\dagger  i \overleftrightarrow{\partial_\mu} X^\nu) 
 \, (\times) $ & 
$\calO_{\ell X4}^{{\tt A},ji}$ &
$(\bar{\ell}_j\gamma^\mu\gamma_5\ell_i )(X_\nu^\dagger  i \overleftrightarrow{\partial_\mu} X^\nu) \, (\times)$
 \\\hline
$ \calO_{\ell X5}^{{\tt V},ji} $ &
$(\bar{\ell}_j\gamma_\mu\ell_i )i\partial_\nu (X^{\mu \dagger} X^\nu - X^{\nu \dagger} X^\mu) 
\,(\times)$ &
$\calO_{\ell X5}^{{\tt A},ji}$ &
$(\bar{\ell}_j\gamma_\mu \gamma_5\ell_i )i \partial_\nu (X^{\mu \dagger} X^\nu - X^{\nu \dagger} X^\mu)\, (\times) $
\\\hline 
$\calO_{\ell X6}^{{\tt V},ji} $ &
$(\bar{\ell}_j\gamma_\mu\ell_i ) i \partial_\nu ( X^\dagger_\rho X_\sigma )\epsilon^{\mu\nu\rho\sigma}\, (\times)$ & 
$ \calO_{\ell X 6}^{{\tt A},ji} $ &
$ (\bar{\ell}_j\gamma_\mu\gamma_5\ell_i )i \partial_\nu (  X^\dagger_\rho X_\sigma)\epsilon^{\mu\nu\rho\sigma} \, (\times)$
\\\hline
\multicolumn{4}{|c|}{\bf Vector DM case B}
\\\hline
$\tilde \calO_{\ell X1}^{{\tt S},ji}$ & 
$(\bar{\ell}_j\ell_i )X_{\mu\nu}^\dagger  X^{\mu\nu}$ &
$\tilde \calO_{\ell X2}^{{\tt S},ji}$ & 
$(\bar{\ell}_j\ell_i )X_{\mu\nu}^\dagger \tilde X^{ \mu\nu}$
\\\hline
$\tilde \calO_{\ell X1}^{{\tt P},ji}$ & 
$(\bar{\ell}_ji\gamma_5\ell_i)X_{\mu\nu}^\dagger X^{ \mu\nu}$ &
$\tilde \calO_{\ell X2}^{{\tt P},ji}$ & 
$(\bar{\ell}_ji \gamma_5\ell_i )X_{\mu\nu}^\dagger \tilde X^{ \mu\nu}$
\\\hline
~$\tilde \calO_{\ell X1}^{{\tt T},ji}$~ & 
~${i \over 2}  ( \bar{\ell}_j \sigma^{\mu\nu}\ell_i )
(X^{\dagger}_{ \mu\rho} X^{\rho}_{\,\nu}-X^{\dagger}_{ \nu\rho} X^{\rho}_{\,\mu}) \, (\times) $~ & 
~$ \tilde \calO_{\ell X2}^{{\tt T},ji}$~ & 
~${1\over 2}  (\bar{\ell}_j \sigma^{\mu\nu}\gamma_5 \ell_i )
(X^{\dagger}_{ \mu\rho} X^{\rho}_{\,\nu}-X^{\dagger}_{ \nu\rho} X^{\rho}_{\,\mu}) \,(\times) $~
\\
\hline
\end{tabular}
 } 
\caption{The leading-order effective operators containing a lepton bilinear and a pair of DM particles in the DSEFT framework.} 
\label{tab:operators}
\end{table}

We focus on the low energy EFT framework extended with DM particles, referred to as the DSEFT~\cite{He:2022ljo,Liang:2023yta}. 
This framework has recently been employed to systematically investigate the DM-electron and DM-nucleus scattering in DM direct detection experiments \cite{Liang:2024lkk,Liang:2024ecw,Liang:2025kkl}.
In the DSEFT, the unbroken SM symmetry $\rm SU(3)_c\times U(1)_{\rm em}$ is imposed on the effective interactions. The relevant degrees of freedom in our current question include the charged leptons $(e,\mu,\tau)$ and a pair of DM fields.
These interactions can be classified according to the spin of the
DM particle. We concentrate on the three commonly studied cases of spin 0, 1/2, and 1, which correspond to scalar, fermion, and vector DM fields, respectively.
We denote the scalar DM as $\phi$ (either complex or real scalar), 
the fermion DM particle as $\chi$ (either Dirac or Majorana fermion), and
the vector DM as $X$ (either complex or real vector), respectively. 
Following the convention in~\cite{He:2022ljo}, we collect the independent leading-order operators involving a LFV charged lepton current and a pair of DM fields in \cref{tab:operators} for the three scenarios. 
Additionally, in the case of the vector DM, we consider two parametrizations based on whether the vector field is represented by the four-vector potential $X_\mu$ (case A) or the field strength tensor $X_{\mu\nu}\equiv\partial_\mu X_\nu - \partial_\nu X_\mu$ (case B).
In the table, the cross symbol `$\times$' indicates the associated operator vanishes when the DM field is real ({\it i.e.}, in the cases of a real scalar, a Majorana fermion, or a real vector). 
The three flavor violating combinations relevant to this work are $(ji)=\{e\mu, e\tau,\mu\tau\}$.  

From \cref{tab:operators}, it is evident that there are two dim-5 and two dim-6 operators for the scalar DM case at leading order. 
In the fermion DM case, there are ten dim-6 operators for each flavor 
combination.  
For LFV vector DM interactions, there are four independent dim-5 operators and twelve dim-6 operators in case A; while in case B, 
only six operators appear at the leading dim-7 order. 
For the two operators $\calO_{\ell X1}^{\tt V(A)}$ in the vector DM case A, 
$\gamma_{(\mu} i \overleftrightarrow{D_{\nu)} } \equiv  \gamma_{\mu} i \overleftrightarrow{D_{\nu} } + \mu \leftrightarrow \nu $,  
where the covariant derivative is denoted as,  $A\overleftrightarrow{D_{\mu} }B \equiv A (D_{\mu}B)-(D_{\mu}A)B$. The dual field strength tensor is defined by $\tilde X^{\mu\nu} = (1/2)\epsilon^{\mu\nu\rho\sigma} X_{\rho\sigma}$.
In the vector DM case A, the decay rates from the insertion of certain operators become divergent in the limit of massless DM particles.
This divergence arises from the longitudinal
component in the polarization sum. For our numerical
analysis, we require the Wilson coefficients (WCs) for these operators to contain an explicit factor
of the DM mass to the minimum power necessary to cancel potential divergences as $m\to 0$ as treated in \cite{He:2022ljo}.

The DSEFT framework is particularly versatile and well-suited 
for processes occurring below ${\cal O}(10~\rm GeV)$ because all heavy particles with masses exceeding this value are integrated out, resulting in a significantly smaller number of active degrees of freedom. 
Conversely, when the full SM gauge symmetry is applied, 
the DSEFT framework can be extended to a SMEFT-like framework. 
In this framework, effective operators are constructed from the DM, the SM chiral fermion, and the Higgs doublet fields, classified under the $\rm SU(3)_c\times SU(2)_L\times U(1)_Y$ symmetries \cite{Brod:2017bsw,Criado:2021trs,Aebischer:2022wnl}. 
It is evident that, when considering operators involving dark particles and SM fermions at the same canonical dimension, the SMEFT-like framework is more restrictive than the DSEFT framework since there are fewer operators in the former due to its larger symmetry requirements. 

For the LFV DSEFT interactions listed in \cref{tab:operators}, there are numerous potential UV models that can generate a subset of these interactions. 
First, in the scalar or fermion DM cases, several well-known models can produce such LFV interactions.
For instance, the scotogenic-like neutrino mass models \cite{Tao:1996vb,Ma:2006km} and their cousins can provide both types of DM candidates while simultaneously yielding LFV interactions at low energy by integrating out the heavy particles. 
For the vector DM LFV interactions, no existing models have been identified in the literature, and we will defer their UV completion to future work.

\section{Three-body LFV decays $\ell_i \to \ell_j +\tt{ DM+DM}$}
\label{sec:3Bdecay}

\begin{figure}[b]
\centering
\includegraphics[width=0.47\textwidth]{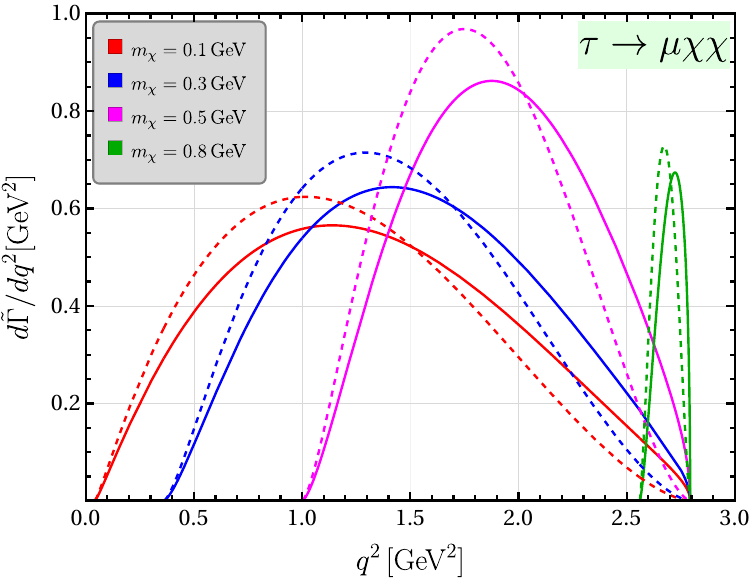}\quad
\includegraphics[width=0.47\textwidth]{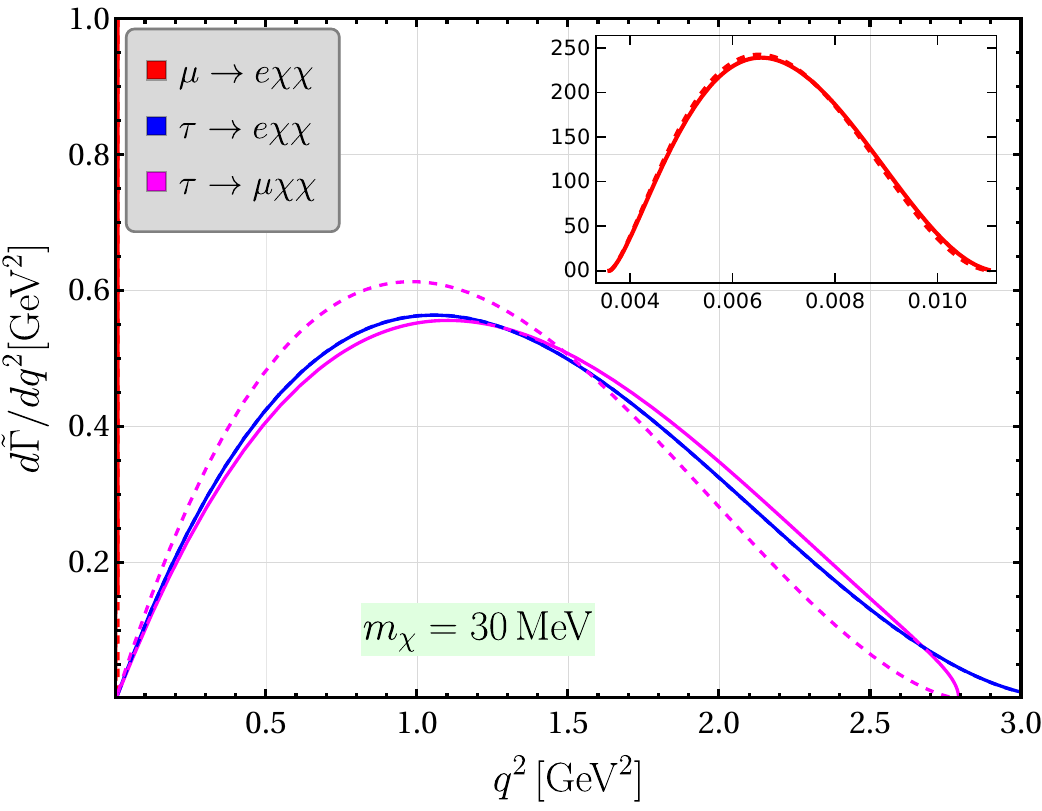}
\caption{Left: Variation of normalized differential decay rate for the process $\tau \to \mu \chi \chi$ considering the operators $\mathcal{O}^{{\tt S}, \mu \tau}_{\ell \chi 1}$ (solid) and $\mathcal{O}^{{\tt P}, \mu \tau}_{\ell \chi 1}$ (dashed) with different DM masses. 
Note that the distribution for $m_{\chi}=0.8$ MeV is scaled by 0.1 along $y$-axis for better visibility.
Right: Variation of normalized differential decay rate for a fixed DM mass, comparing different processes for operators $\mathcal{O}^{{\tt S}, ji}_{\ell \chi1}$ (solid) and $\mathcal{O}^{{\tt P}, ji}_{\ell \chi1}$ (dashed). The inset is for $\mu\to e\chi\chi$ due to much smaller $q^2$. 
}
\label{fig:diff.dist.mass}
\end{figure}

\subsection{The $q^2$-distribution}
\label{sec:qsqrd_dist}

Experimentally, the $q^2$-distribution for a three-body decay can provide important insights into the underlying interactions,
particularly for decay modes that involve invisible particles. 
Therefore, it is worthwhile to explore the distribution behavior in our cases. 
In the following part, we analyze the $q^2$-distribution
for the cLFV decays involving a pair of DM particles, $\ell_i(p) \to \ell_j(k) +{\tt DM}(k_1)+{\tt DM}(k_2)$. 
Here, $q^2\equiv(p-k)^2 = (k_1+k_2)^2$ is defined as the invariant mass squared of the invisible particles,
which may be measured through the momentum transfer of the charged leptons. 
The matrix elements and the differential rate expressions for the decay modes in the four DM scenarios are provided in \cref{app:Mi2jDMDM} and \cref{app:diff.rate}, respectively.

\begin{figure}[b]
\centering
\includegraphics[width=0.47\textwidth]{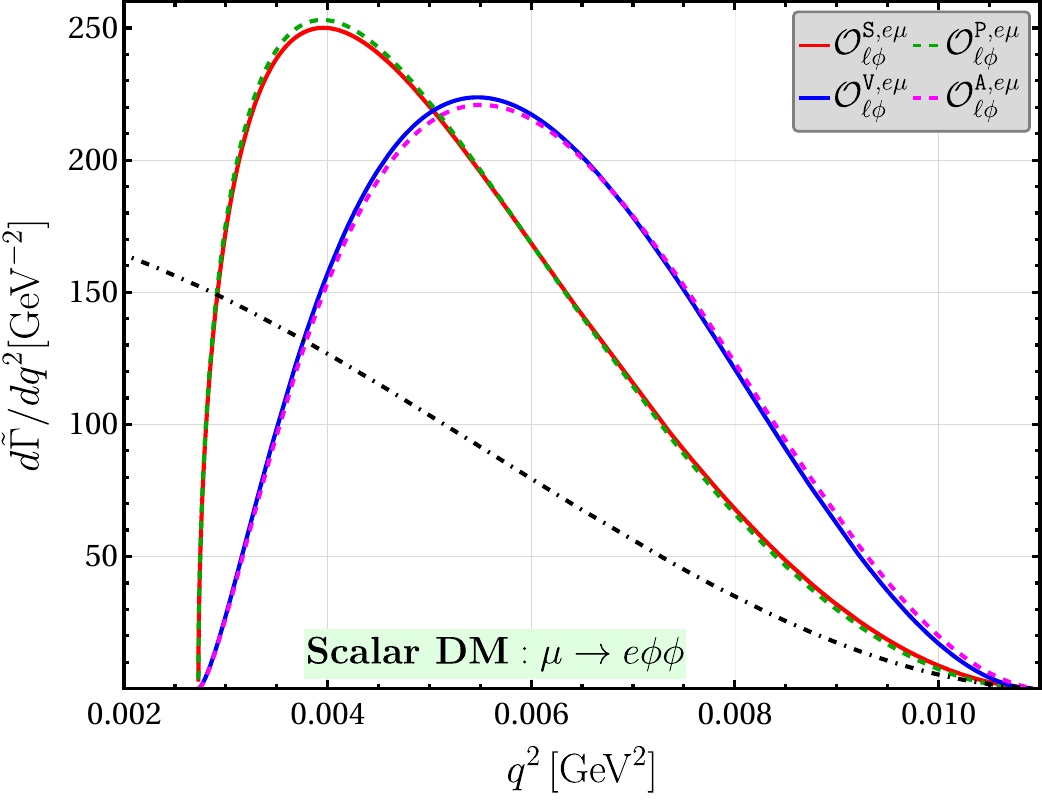}\quad
\includegraphics[width=0.47\textwidth]{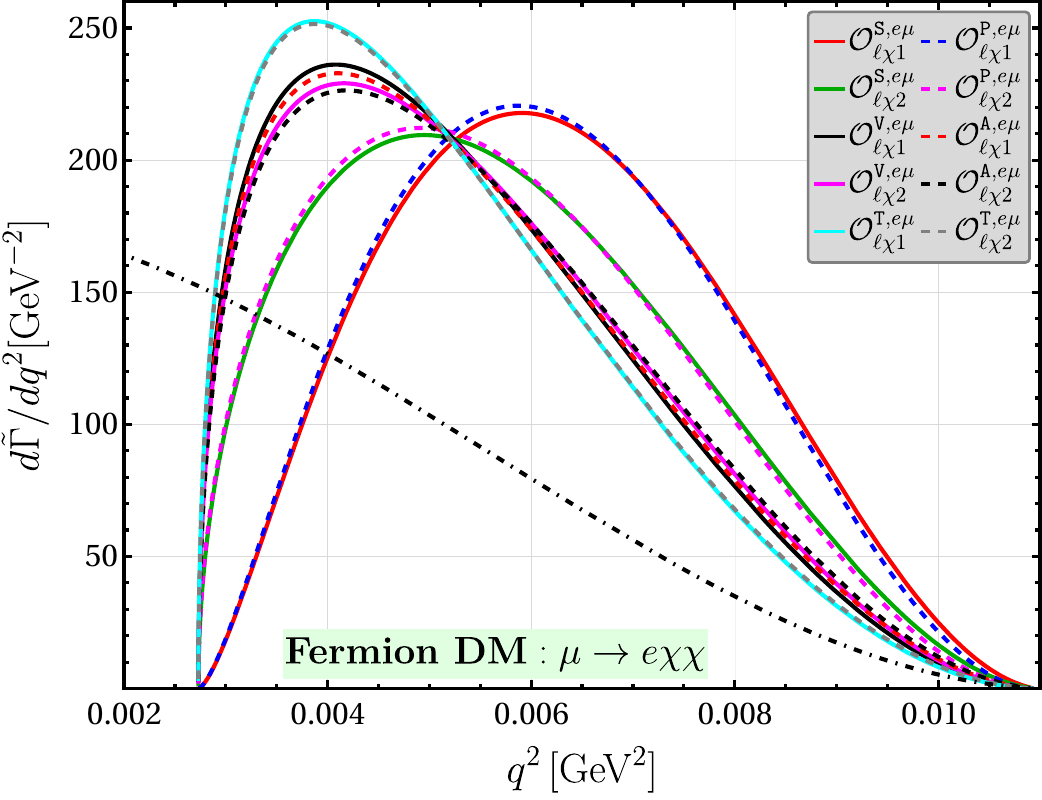}
\\
\includegraphics[width=0.47\textwidth]{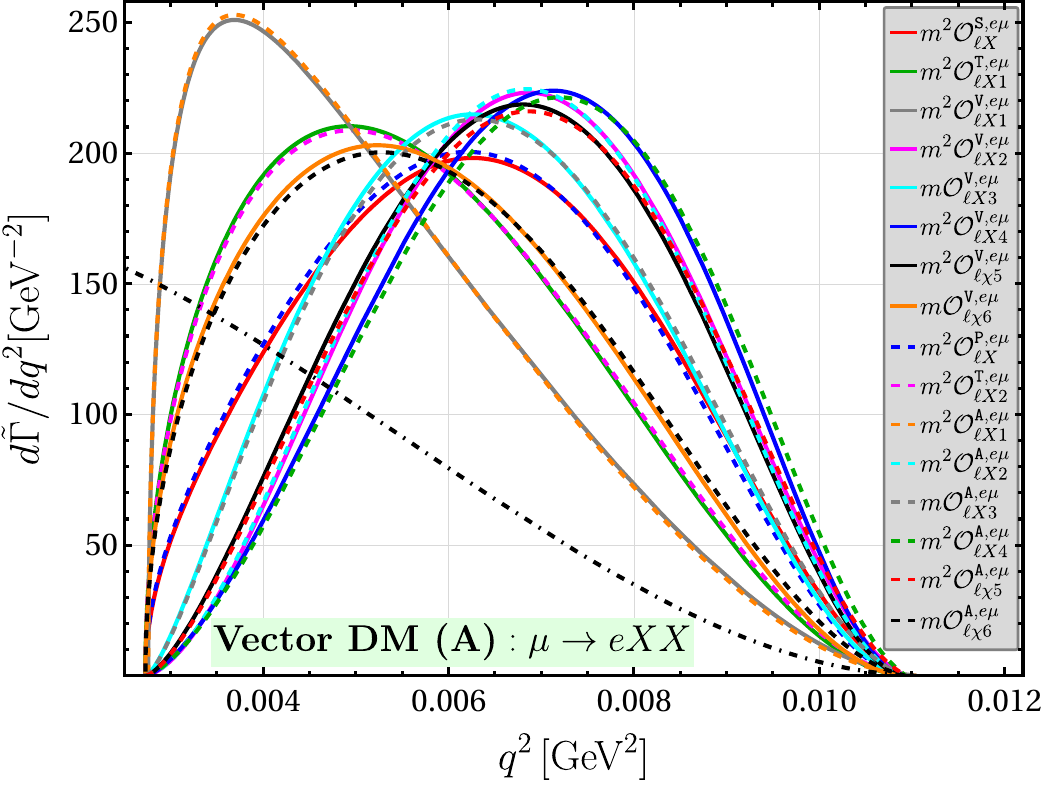}\quad
\includegraphics[width=0.47\textwidth]{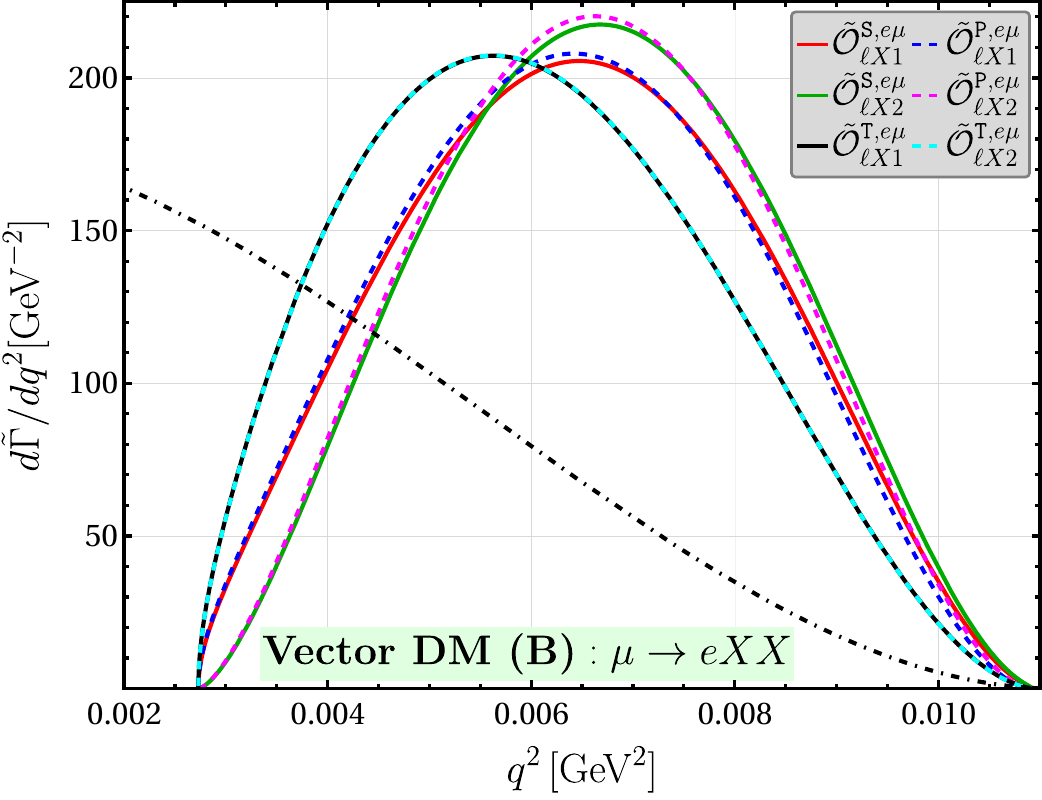}
\caption{Differential decay rate distribution for the process $\mu \to e +{\tt DM+DM}$ with a fixed DM mass of 26~MeV, from different DSEFT operators. The darker dot-dashed black line represents the differential rate for the SM process $\mu^- \to e^- \bar{\nu}_e \nu_{\mu}$,
which contributes at leading order.
}
\label{fig:diff.dist.mu2e}
\end{figure}

\begin{figure}[t]
\centering
\includegraphics[width=0.47\textwidth]{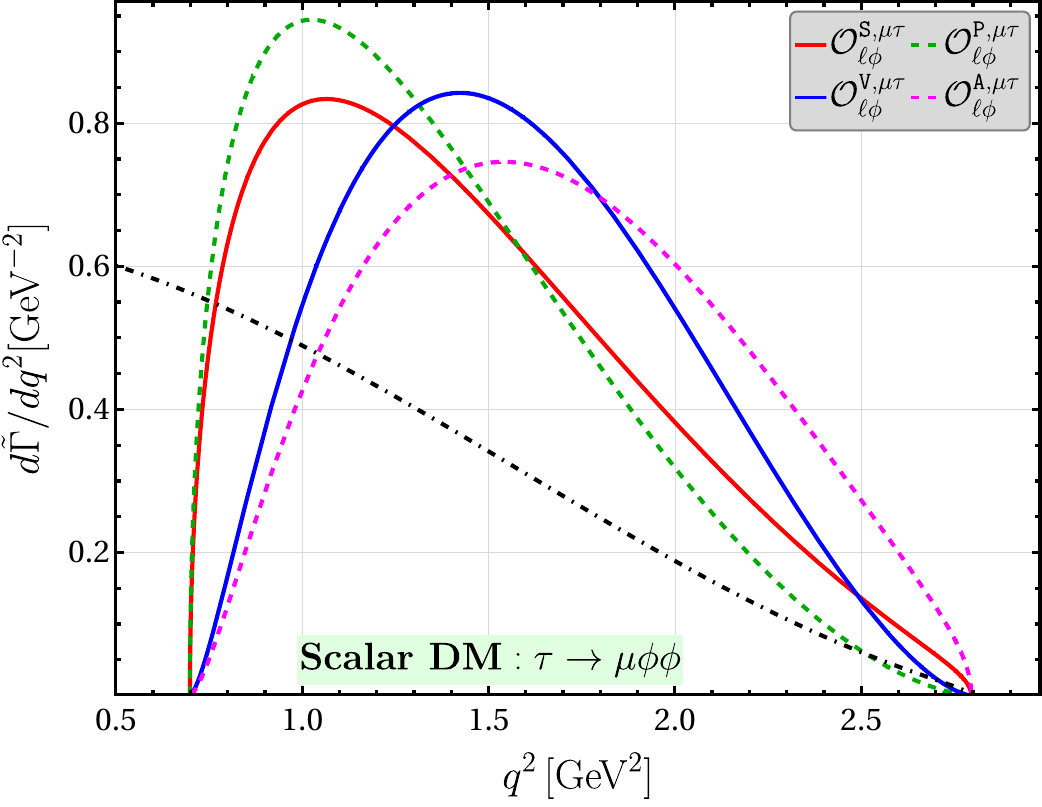}~
\includegraphics[width=0.47\textwidth]{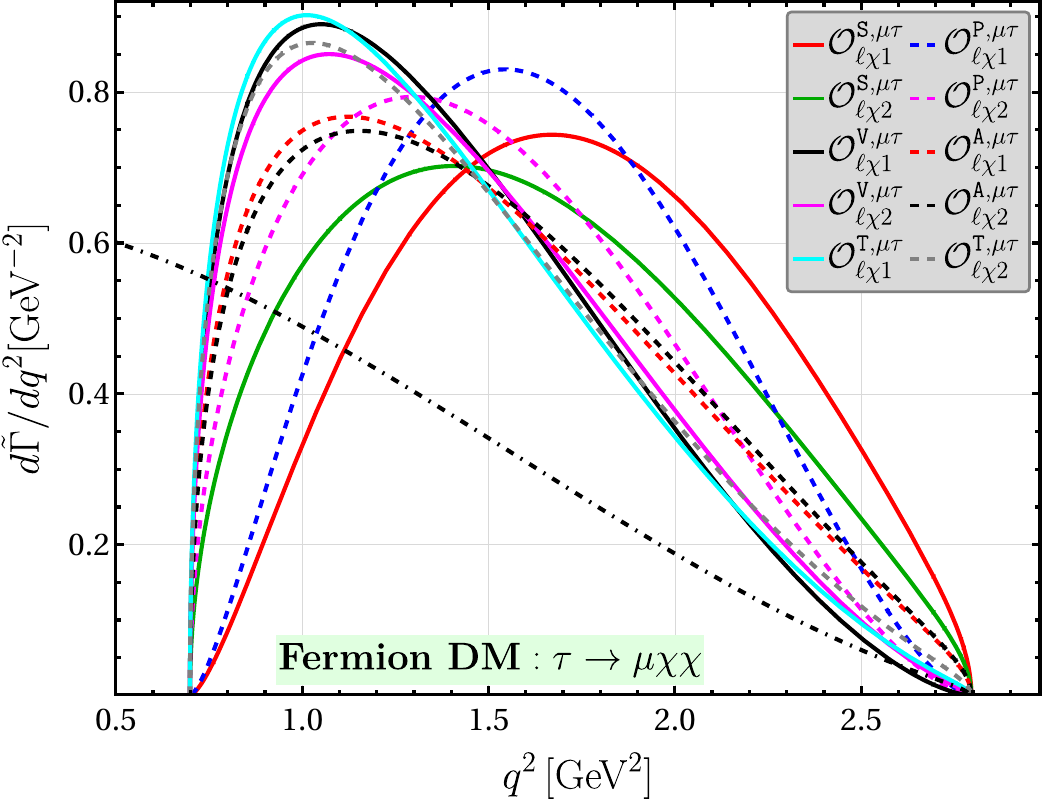}
\\
\includegraphics[width=0.47\textwidth]{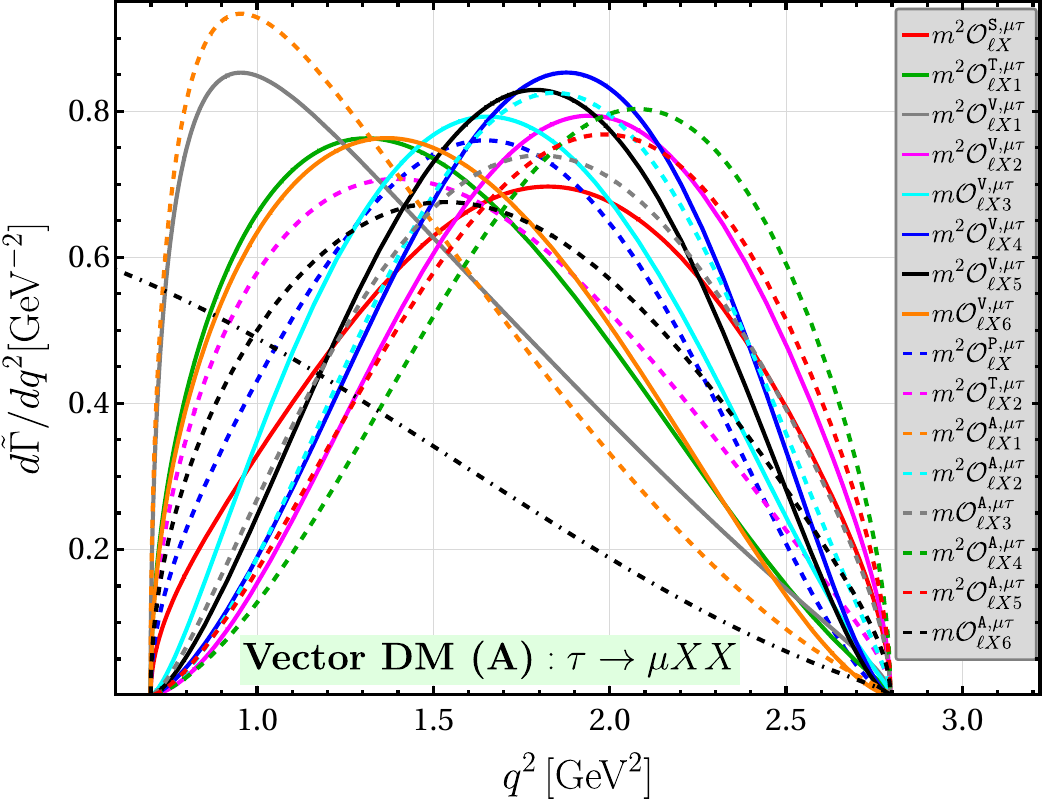}~
\includegraphics[width=0.47\textwidth]{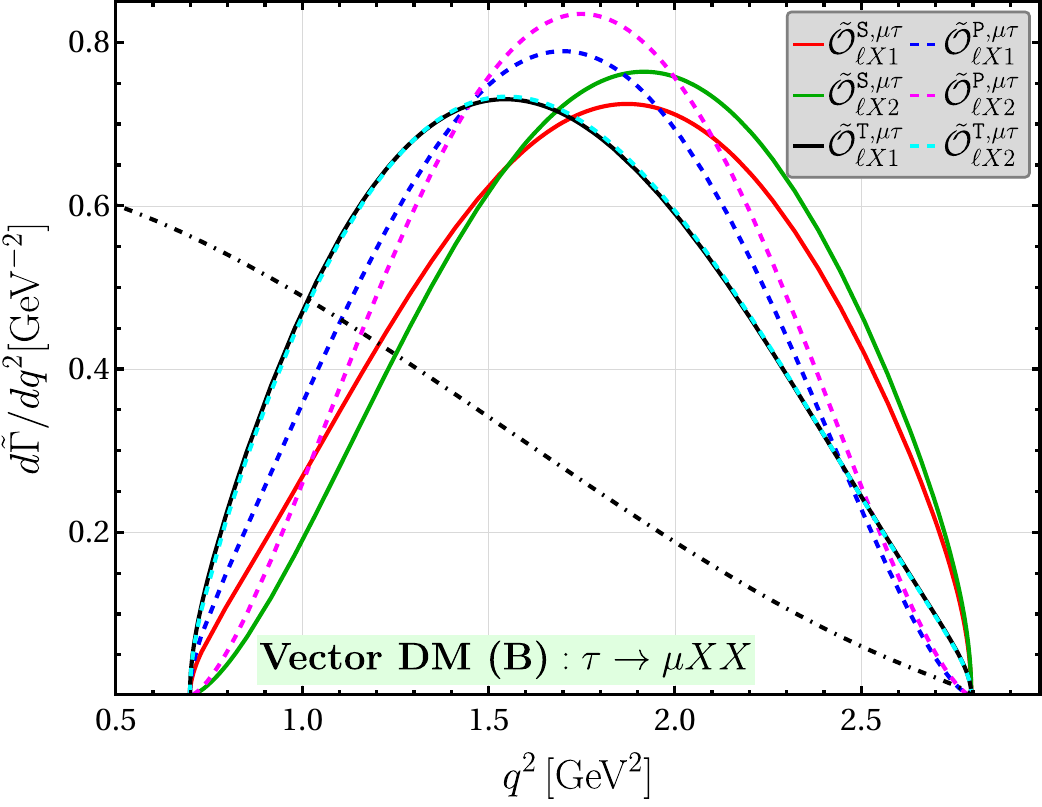}
\caption{Same as \cref{fig:diff.dist.mu2e} but for the process $\tau \to \mu + {\tt DM + DM}$ with a DM mass of 418 MeV. The darker dot-dashed black line represents the decay rate for the SM process $\tau^- \to \mu^- \bar{\nu}_{\mu} \nu_{\tau}$.}
\label{fig:diff.dist.tau2mu}
\end{figure}

The normalized differential decay rate, $d\tilde\Gamma/d q^2 \equiv (1/\Gamma)(d\Gamma/dq^2)$,
for the process $\tau \to \mu \chi\chi$ is shown in the left panel of \cref{fig:diff.dist.mass}  for varying DM mass considering the  operators $\mathcal{O}^{{\tt S}, \mu \tau}_{\ell \chi 1}$ and $\mathcal{O}^{{\tt P}, \mu \tau}_{\ell \chi 1}$ in the case of fermion DM. As the DM mass increases, the available phase space for its production becomes increasingly restricted.
From these two representative operators, we see that the $q^2$-distribution is influenced by the DM mass, the mass splitting between the initial- and final-state leptons, and the Lorentz structures of the SM leptonic current and DM current of the effective operators (see \cref{fig:diff.dist.mu2e,fig:diff.dist.tau2mu} and Eqs.~\eqref{eq:dGammadq2_S}-\eqref{eq:dGammadq2_VB}). 
The right panel of \cref{fig:diff.dist.mass} illustrates the variation in the differential distributions for the three processes with different lepton mass splittings, keeping the DM mass fixed. 
For the process $\tau \to e \chi\chi$, the distributions of the operators with leptonic scalar ($\mathcal{O}^{{\tt S}, e\tau}_{\ell \chi1}$) and pseudoscalar ($\mathcal{O}^{{\tt P}, e\tau}_{\ell \chi1}$) structures
overlap significantly (shown in blue), due to the electron mass being negligible in comparison to $m_\tau$. 
In contrast, for $\tau \to \mu \chi \chi$, the distributions remain distinguishable (shown in magenta). A similar behavior is noticed for the DSEFT operators with SM vector and axial-vector currents, as well as tensor and pseudotensor currents. From the left panel of \cref{fig:diff.dist.mass}, we infer that the $q^2$-distribution is an important observable that can be used to obtain mass information about the DM and differentiate between the operator structures.

\cref{fig:diff.dist.mu2e,fig:diff.dist.tau2mu} present the differential distributions for the processes $\mu \to e+{\tt DM+DM}$ and $\tau \to \mu +{\tt DM+DM}$, respectively, from the insertion of various DSEFT operators listed in \cref{tab:operators} for all possible DM scenarios.\footnote{We do not present the results for $\tau \to e+{\tt DM+DM}$ since they are entirely analogous to those of the $e\mu$ flavor case, where the effect of the electron mass is negligible.}
The corresponding SM counterparts ($\mu^-\to e^-\bar \nu_e \nu_\mu$ and $\tau^-\to \mu^-\bar \nu_\mu \nu_\tau$) are represented by the darker dot-dashed curves for comparison.
Please note that we have included mass factors to appropriate powers for the operators in the vector DM case A to cancel the potential divergence in the limit of $m\to 0$.
The DM mass is taken to be a quarter of the lepton mass splitting, i.e., $m=\Delta m_{ij}/4$ with $\Delta m_{ij}\equiv m_i - m_j$. 
These distributions are distinguishable for operators with different Lorentz structures, such as scalar ({\tt S}), vector ({\tt V}), and tensor ({\tt T1}) leptonic currents. However, due to the negligible electron mass, it is clear from \cref{fig:diff.dist.mu2e} that the differential distributions of the process $\mu \to e +{\tt DM +DM}$ are unable to resolve the operators with {\tt S/V/T1} lepton currents from the pseudoscalar ({\tt P})/axial-vector ({\tt A})/pseudotensor ({\tt T2}) currents (for a fixed DM current) for all DM scenarios. On the other hand, for $\tau \to \mu +{\tt DM +DM}$ in \cref{fig:diff.dist.tau2mu}, we see that in the case of the scalar and fermion DM scenarios, the distributions are clearly distinguishable between {\tt S/P/T1} operators and {\tt P/A/T2} operators for a fixed DM current. 
A similar behavior is evident in the vector DM case A. Notably, for the operators $\mathcal{O}^{\tt V, \mu \tau}_{\ell X 1-6}$ with the same vector leptonic current, the different Lorentz structures in the DM current lead to distinct distributions for these operators. For vector DM case B, the differential distributions of the two scalar leptonic current operators $\tilde{\cal{O}}^{\tt S, \mu \tau}_{\ell X 1(2)}$ are distinguishable from those of the pseudoscalar current operators $\tilde{\cal{O}}^{\tt P, \mu \tau}_{\ell X 1(2)}$, whereas the distributions of the tensor and pseudotensor operators $\tilde{\cal{O}}^{\tt T, \mu \tau}_{\ell X 1(2)}$ remain indistinguishable. 

However, from an experimental perspective, a quantitative discrimination among different operators is more challenging. This difficulty arises because it requires observing an excess of events beyond the SM prediction, performing a pseudo–rest frame reconstruction of the tau lepton~\cite{ARGUS:1995bjh}, as well as incorporating additional experimental inputs as detailed in \cite{Belle-II:2022heu,Belle:2025bpu}.

In the following subsections, we will impose constraints on the effective scale associated with the WCs of the corresponding operators. In the scalar, fermion, and vector B DM cases, the effective scale is defined through the WC via $C_d\equiv 1/\Lambda_{\rm eff}^{d-4}$ for operators with dimensions $d=5,6,7$.
However, in the vector DM case A, the differential decay widths (and total decay widths) diverge as the DM mass approaches zero, $m_X \to 0$. This divergence is a known feature of vector DM described in terms of a four-vector potential. To circumvent this issue and establish meaningful constraints for smaller values of $m_X$, we posit that each relevant WC exhibits a dependence on the DM mass, with the power of $m_X$ determined by the number of independent vector four-potentials that cannot be reduced to field strength tensors. Thus, we define the effective scale associated with each operator in the following way:
\begin{align}
C_{\ell X}^{\tt S,P} \equiv \frac{m_X^2}{\Lambda_{\rm eff}^3}, \qquad 
C_{\ell X1,2}^{\tt T} \equiv \frac{m_X^2}{\Lambda_{\rm eff}^3}, \qquad 
C_{\ell X1,2,4,5}^{\tt V,A} \equiv \frac{m_X^2}{\Lambda_{\rm eff}^4}, \qquad 
C_{\ell X3,6}^{\tt V,A} \equiv \frac{m_X}{\Lambda_{\rm eff}^3}.
\label{eq:norm.couplings}
\end{align}

\subsection{Constraints from $\mu^- \to e^-\nu_{e} \bar{\nu}_{\mu}$}

\begin{figure}
\centering
\includegraphics[height=5cm, width=7.cm]{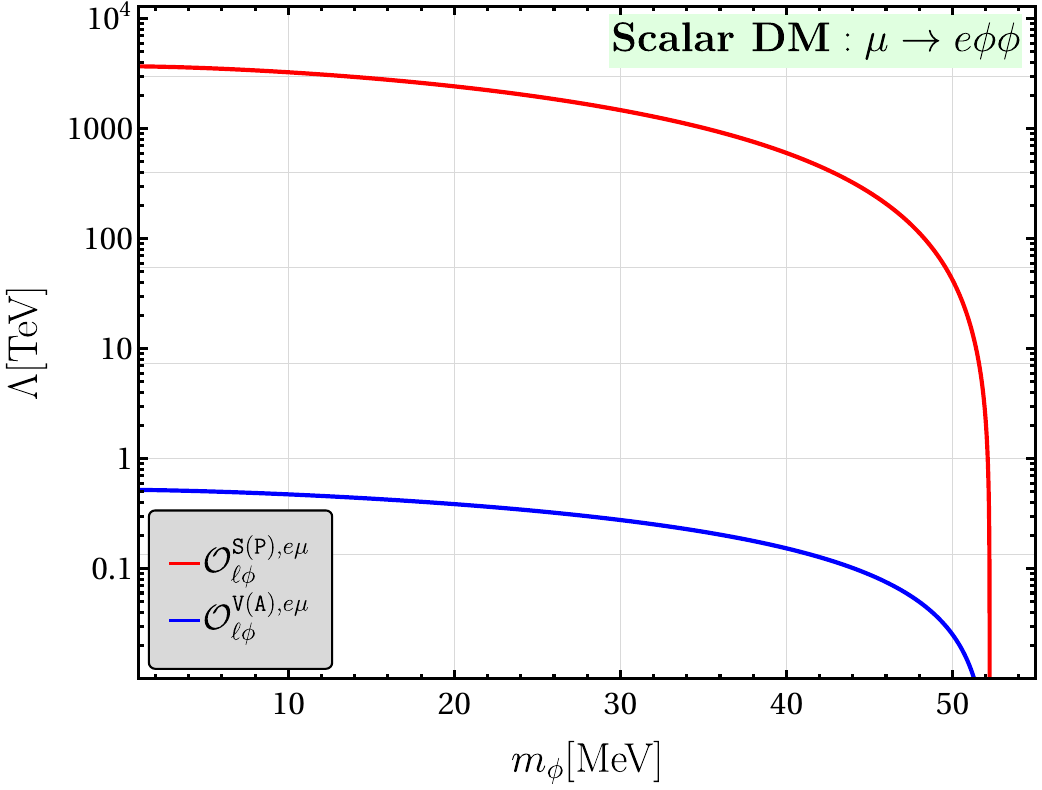}\quad
\includegraphics[height=5cm, width=7.cm]{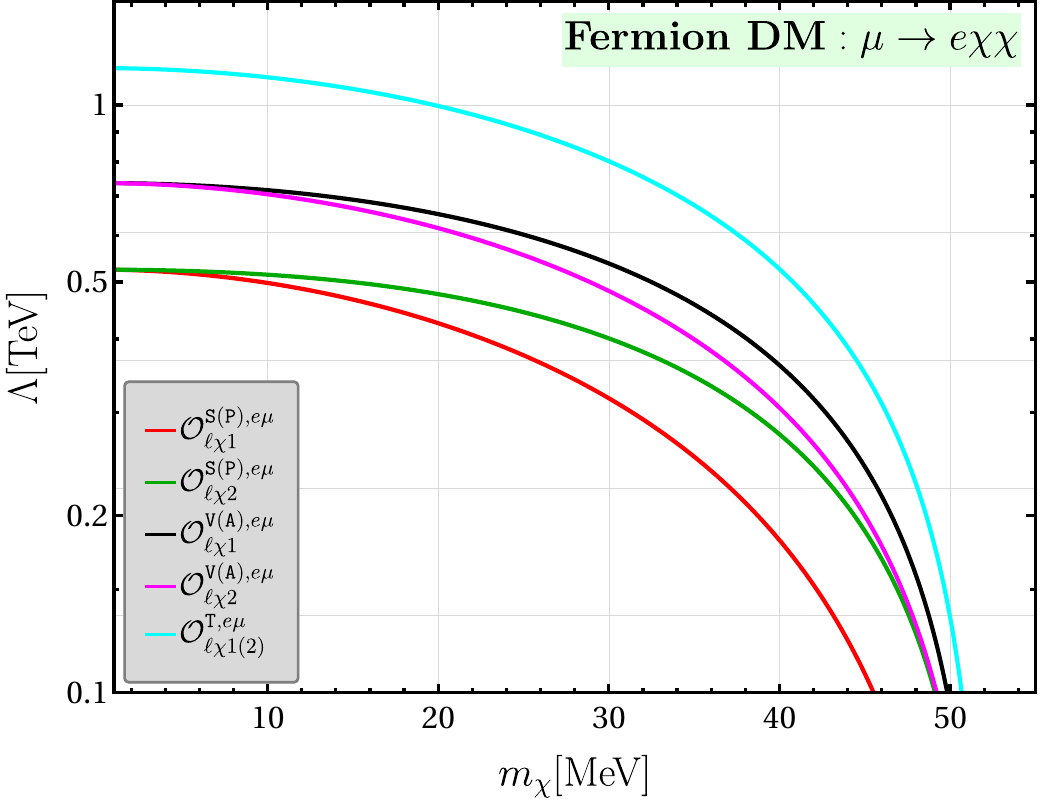}
\vspace{0.5em}
\\
\includegraphics[height=5cm, width=7.cm]{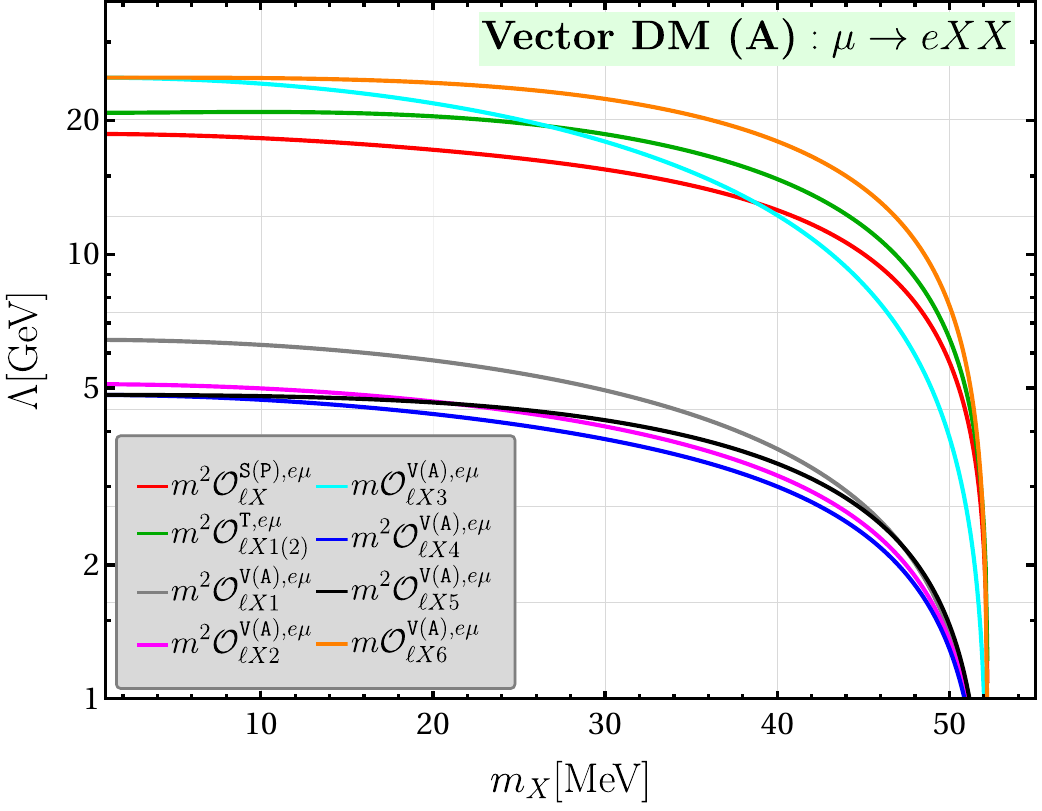}\quad
\includegraphics[height=5cm, width=7.cm]{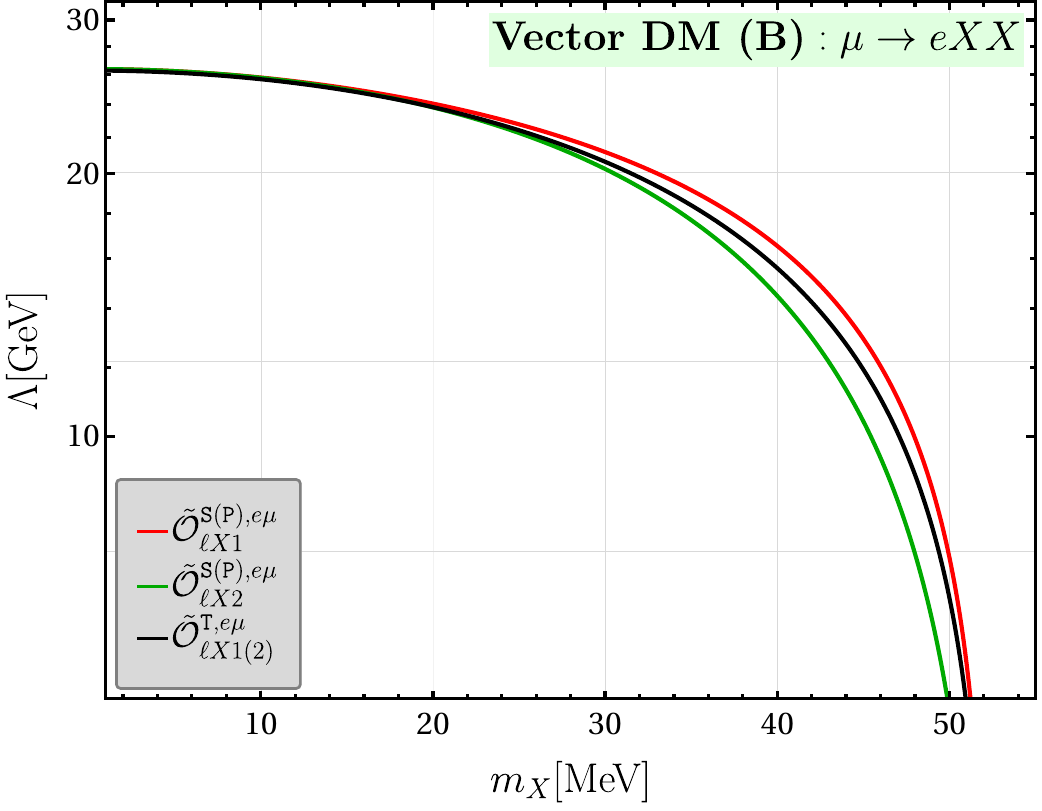}
\caption{Constraints on the effective scale $\Lambda$ as a function of the DM mass $m$ for different sets of DSEFT operators from the process $\mu \to e + {\tt DM+DM}$.}
\label{fig:const.mu2e}
\end{figure}

Since the neutrino pair is invisible to the detector, we use the upper bound on the LFV muon decay $\mu^- \to e^-\nu_{e} \bar{\nu}_{\mu}$ to set conservative limits. Specifically, we require that ${\cal B}(\mu\to e +{\tt DM+DM})\lesssim 0.012$.\footnote{More stringent bounds are expected to be obtained by fitting the Michel parameters, as they have been measured with high precision. However, such a dedicated analysis is beyond the scope of the present work and will be deferred to future studies.} The experimental constraints in the $\Lambda-m$ plane, derived using Eqs.~\eqref{eq:dGammadq2_S}-\eqref{eq:dGammadq2_VB}, are presented in \cref{fig:const.mu2e}. These constraints encompass all possible DM scenarios arising from $\mu \to e +{\tt DM+DM}$ decay.
The constraints on the pair of operators that share the same DM current but differ in the lepton current by a $\gamma_5$ are nearly identical in the $\Lambda-m$ plane, owing to the negligible electron mass. Therefore, we represent each pair with a single curve in the plot.
We note that, throughout our analysis, we present the constraints on $\Lambda$ for complex scalar/vector and Dirac fermion DM. For real scalar/vector and Majorana fermion DM, the constraints on $\Lambda$ are stronger by a factor of $2^{1/[2(d-4)]}$, where $d$ denotes the dimension of the effective operator (including the mass factors shown in the legends for the vector DM case A).

{\bf Scalar DM case}: 
In the top-left panel of \cref{fig:const.mu2e}, we present the constraints in the $\Lambda-m_\phi$ plane for the scalar DM scenario. Owing to their lower dimension, leptonic scalar and pseudoscalar current operators $\mathcal{O}^{{\tt (P)S}, e \mu}_{\ell \phi}$ have tighter constraints compared to vector and axial-vector current operators $\mathcal{O}^{{\tt V (A)}, e \mu}_{\ell \phi}$. 
For the operators $\mathcal{O}^{{\tt S(P)}, e \mu}_{\ell \phi}$, when the DM mass is far from $\Delta m_{\mu e}/4$ ($m_\phi \sim 1$ MeV), the constraint on $\Lambda$ is approximately 3700 TeV. At $\Delta m_{\mu e}/4$ ($m_\phi \sim 26$ MeV), the constraint reduces to 1800 TeV. On the other hand, for the operators $\mathcal{O}^{{\tt V(A)}, e \mu}_{\ell \phi}$, the constraint on $\Lambda$ is approximately 520 GeV at $m_\phi \sim 1$ MeV, and it decreases to 300 GeV at $m_\phi \sim 26$ MeV.

{\bf Fermion DM case}: 
The top-right plot of \cref{fig:const.mu2e} illustrates the constraints in the fermion DM case. As illustrated from the plot, the operators $\mathcal{O}^{{\tt T}, e \mu}_{\ell \chi 1(2)}$ ($\mathcal{O}^{{\tt S(P)}, e \mu}_{\ell \chi 1}$) receive the most (least) stringent constraint.  For a DM mass of $ m_\chi \sim 1 $ (26) MeV, the constraint on $\Lambda$ is approximately 1.16 (0.89) TeV for the operators $\mathcal{O}^{{\tt T}, e \mu}_{\ell \chi 1(2)}$. Similarly, for the operators $\mathcal{O}^{{\tt S(P)}, e \mu}_{\ell \chi 1}$, the constraint on $\Lambda$ is approximately 525 (436) GeV for the same DM mass value.

{\bf Vector DM case}: 
The constraints on the vector DM case A are shown in the $\Lambda–m_X$ plane in the bottom-left panel of \cref{fig:const.mu2e}. In this scenario, the operators $\mathcal{O}^{{\tt V(A)}, e \mu}_{\ell X 6}$ have the most stringent constraint on $\Lambda$, while $\mathcal{O}^{{\tt V(A)}, e \mu}_{\ell X 4}$ exhibit the weakest constraint. For a DM mass of $m_X \sim 1$ (26) MeV, the constraint on $\Lambda$ is approximately 25 (23) GeV for the operators $\mathcal{O}^{{\tt V(A)}, e \mu}_{\ell X 6}$. In the case B, all operators exhibit comparable constraints due to the same dimensionality and parameterizations of the effective scale. At a DM mass of $m_X \sim 1$ (26) MeV, the constraint on $\Lambda$ is approximately 26 (22) GeV for all operators.
It should be noted that the relatively weak constraints on the effective scale are due to the higher dimensionality of the operators, such as the parametrization in \cref{eq:norm.couplings} for the case A.

\subsection{Constraints from $\tau \to e+\rm inv.$}

\begin{figure}[t]
\centering
\includegraphics[height=5cm, width=7.cm]{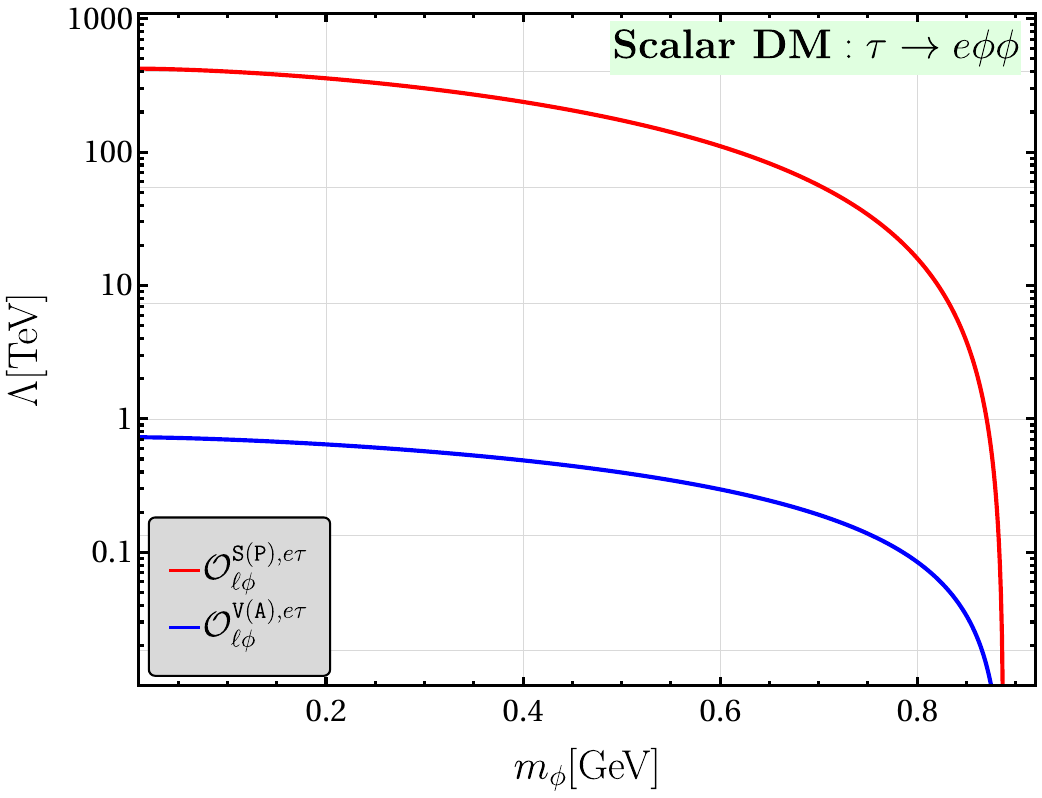}\quad
\includegraphics[height=5cm, width=7.cm]{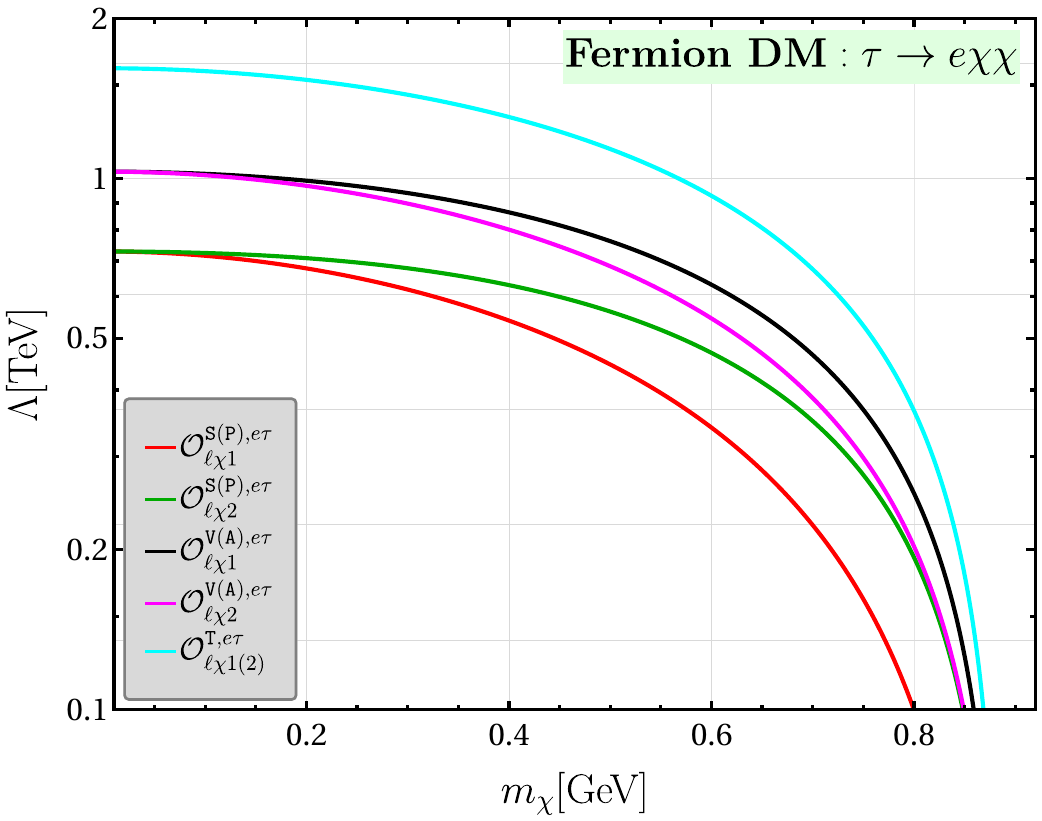}
\vspace{0.5em}
\\
\includegraphics[height=5cm, width=7.cm]{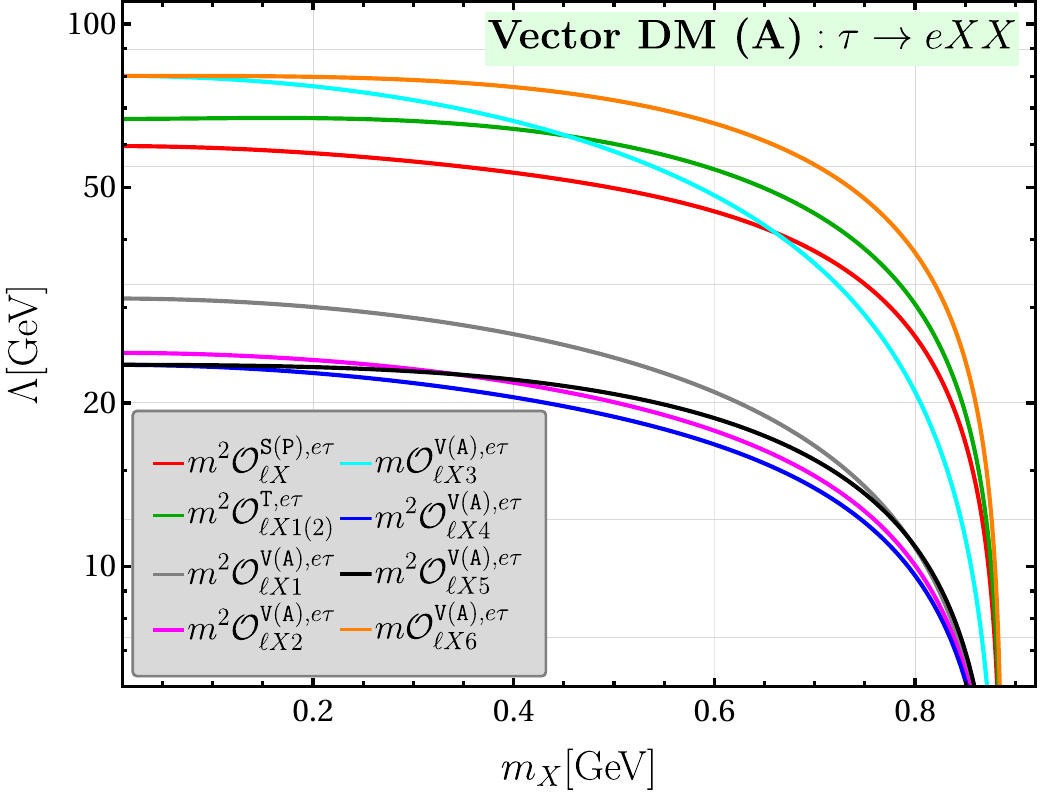}\quad
\includegraphics[height=5cm, width=7.cm]{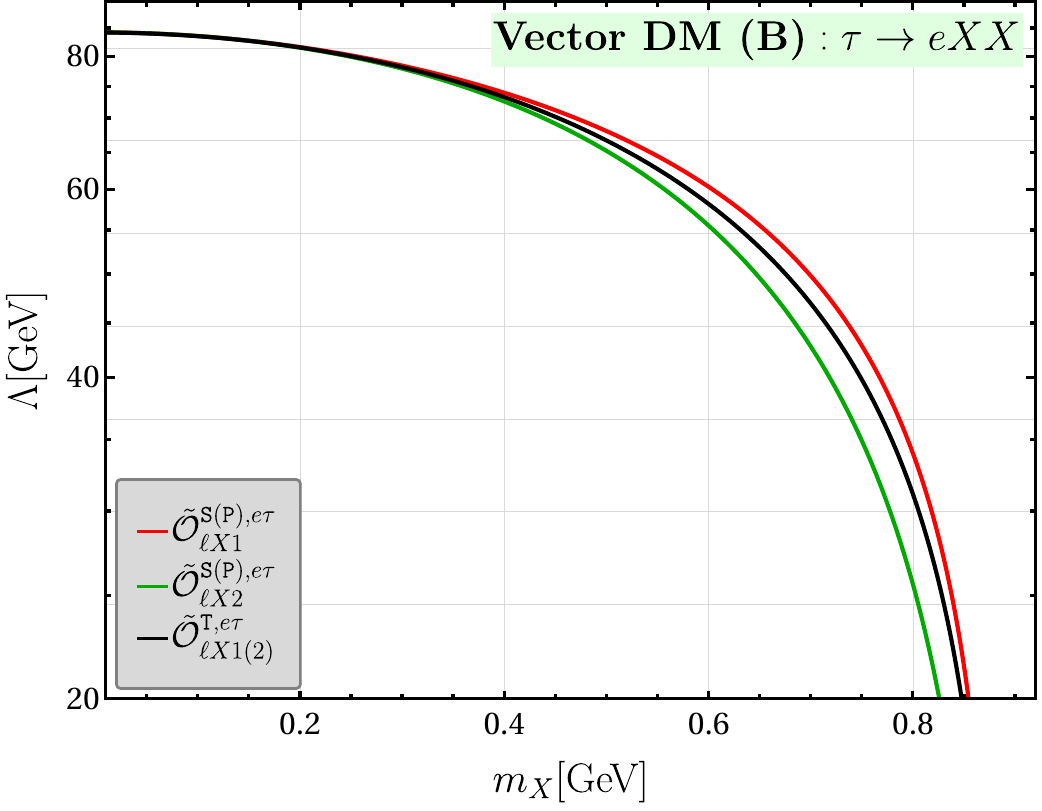}
\caption{Same as \cref{fig:const.mu2e} but for the process $\tau \to e +{\tt DM+DM}$.}
\label{fig:const.tau2e}
\end{figure}

For the $e\tau$ flavor combination, we use the derived bound $\mathcal{B}(\tau \to e+{\tt DM +DM}) < 5.8 \times 10^{-4} $ to set our limits. In \cref{fig:const.tau2e} we depict the constraints in the $\Lambda-m$ plane from the flavor violating $\tau \to e + {\tt DM +DM }$ decay for all the DM scenarios. Similar to the $\mu \to e +{\tt DM+DM}$ decay, for operators with a fixed DM current, the leptonic {\tt S/V/T1} and {\tt P/A/T2} current operators are subject to nearly identical constraints and are therefore represented by the same curves in the $\Lambda-m$ plane.

{\bf Scalar DM case}:
The constraints for the scalar DM scenario are shown in the top-left panel of \cref{fig:const.tau2e}. For the operators $\mathcal{O}^{{\tt S(P)}, e \tau}_{\ell \phi}$, considering the same $m_{\phi}$ and $\Lambda$, the total decay width for $\tau \to e \phi \phi$ is four orders of magnitude stronger than $\mu \to e \phi \phi$. However, the experimental upper bound on the decay width of $\tau \to e \phi \phi$ is six orders of magnitude weaker than $\mu \to e \phi \phi$. As a result, compared to the constraints on $\Lambda$ from the process $\mu \to e\phi\phi$, the operators $\mathcal{O}^{{\tt S(P)}, e \tau}_{\ell \phi}$ receive a constraint that is an order of magnitude weaker. On the other hand, for the operators $\mathcal{O}^{{\tt V(A)}, e \tau}_{\ell \phi}$, considering the same $m_{\phi}$ and $\Lambda$, the total decay width for $\tau \to e \phi \phi$ is six order stronger than $\mu \to e \phi \phi$. Therefore, the constraints on $\mathcal{O}^{{\tt V(A)}, e \tau}_{\ell \phi}$ are of a comparable magnitude as the $e\mu$ flavor case. For the operators $\mathcal{O}^{{\tt S(P)}, e \tau}_{\ell \phi}$, when the DM mass is far from $\Delta m_{\tau e}/4$ ($m_\phi \sim 0.01$ GeV), the constraint on $\Lambda$ is approximately 421 TeV. As the DM mass approaches $\Delta m_{\tau e}/4$ ($m_\phi \sim 0.44$ GeV), the constraint reduces to 209 TeV. In contrast, for the operators $\mathcal{O}^{{\tt V(A)}, e \tau}_{\ell \phi}$, the constraint on $\Lambda$ is approximately 725 (499) GeV for a DM mass of $m_\phi \sim 0.01$ (0.44) GeV.

{\bf Fermion DM case}: 
In comparison to the constraints on $\Lambda$ from $\mu \to e \chi\chi$ decay, a slight improvement is noticed as shown in the top-right panel of \cref{fig:const.tau2e}. For all operators, the $\tau \to e \chi\chi$ decay provides a factor of approximately 1.38 improvement on $\Lambda$ when the DM mass is far below $\Delta m_{\tau e}/4$. For a DM mass of $m_\chi \sim 0.01$ (0.44) GeV, the constraint on $\Lambda$ is around 1.61 (1.23) TeV for the operators $\mathcal{O}^{{\tt T}, e \tau}_{\ell \chi 1(2)}$. Likewise, for the operators $\mathcal{O}^{{\tt S(P)}, e \tau}_{\ell \chi 1}$, the constraint on $\Lambda$ is approximately 726 (497) GeV for the same DM mass.

{\bf Vector DM case}: 
In the case A, depending on specific EFT operators, the constraint on $\Lambda$ is improved by a factor of approximately 3 to 5 compared to the corresponding one with $e \mu$ flavor combination from $\mu \to e XX$ decay. The operators $\mathcal{O}^{{\tt V(A)}, e \tau}_{\ell X 6}$ receive the tightest constraint on $\Lambda$, while the operators $\mathcal{O}^{{\tt V(A)}, e \tau}_{\ell X 4}$, similar to $\mathcal{O}^{{\tt V(A)}, e \mu}_{\ell X 4}$, have the weakest constraint on their associated effective scale. For a DM mass of $m_X \sim 0.01$ (0.44) GeV, the constraint on $\Lambda$ is approximately 80 (74) GeV for the operators $\mathcal{O}^{{\tt V(A)}, e \mu}_{\ell X 6}$. Similarly, for the operators $\mathcal{O}^{{\tt V(A)}, e \mu}_{\ell X 4}$, the corresponding constraint is approximately 23 (20) GeV for the same DM mass. For the vector DM case B, like $\mu \to e X X$ decay, the degeneracy for small $m_X$ still holds among all operators in the $\Lambda-m_X$ plane with a global enhancement of a factor of 3 relative to $e \mu$ case. At a DM mass of $m_X \sim 0.01$ (0.44) GeV, the constraint on $\Lambda$ is approximately 84 (71) GeV for all operators.

\subsection{Constraints from $\tau \to \mu+\rm inv.$}
\begin{figure}[t]
\centering
\includegraphics[height=5cm, width=7.cm]{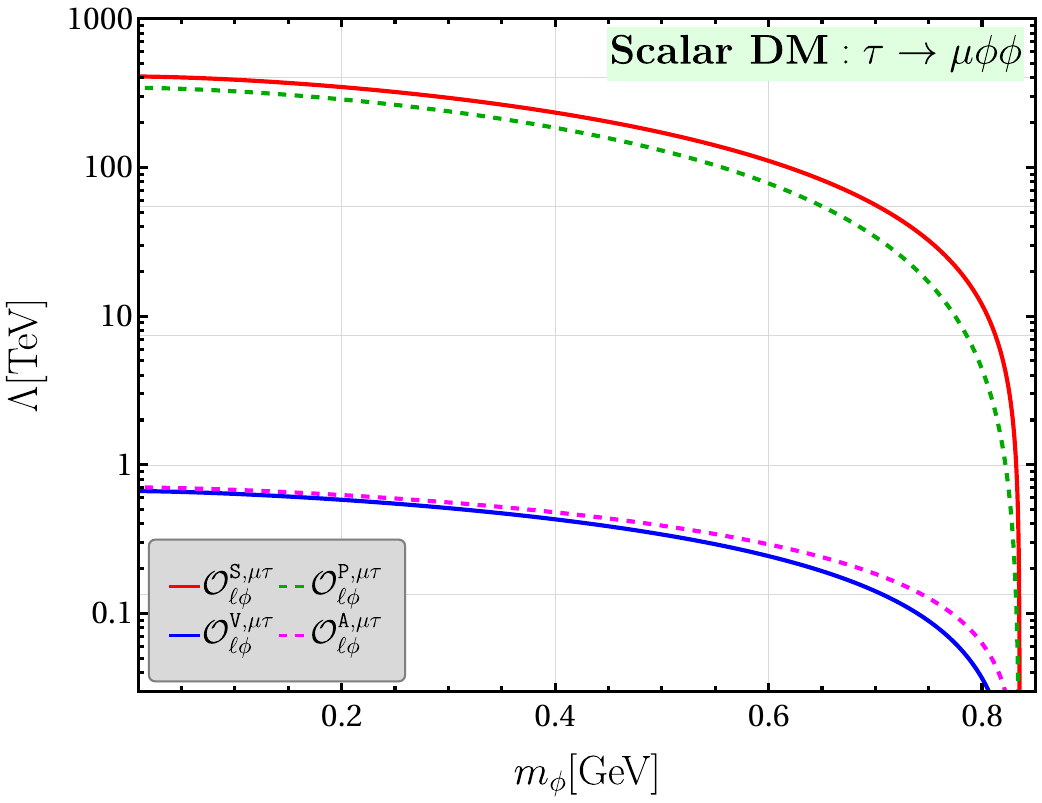}\quad
\includegraphics[height=5cm, width=7.cm]{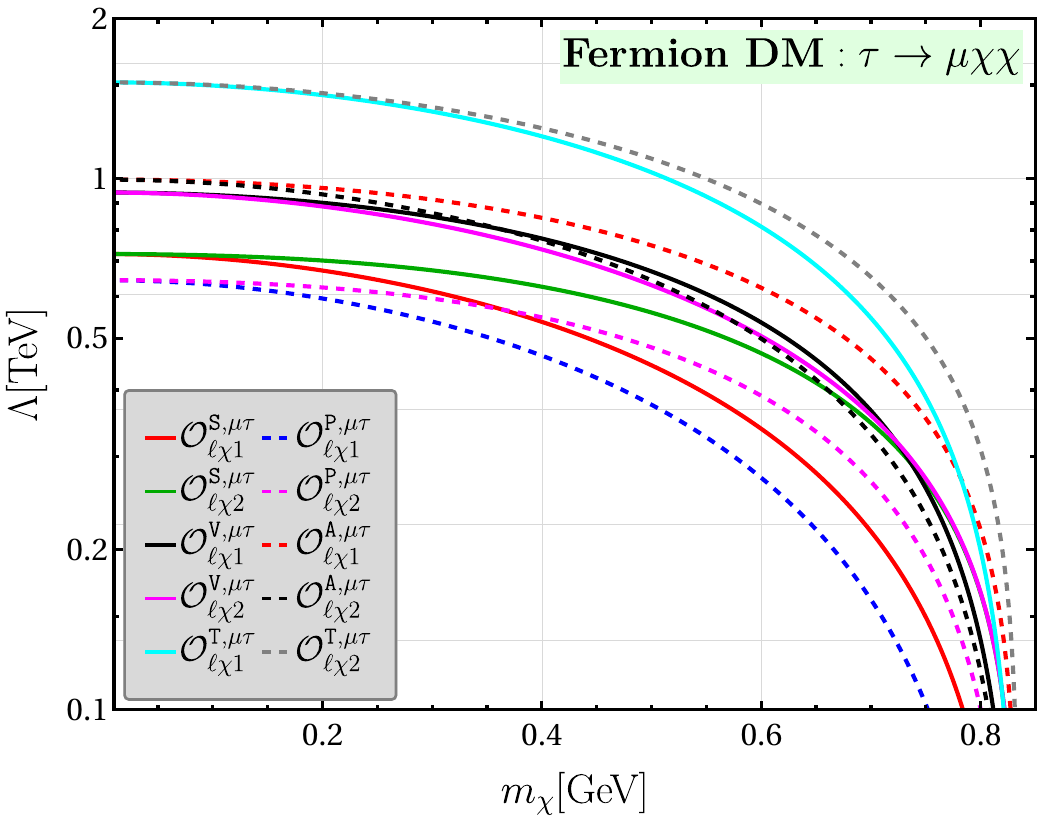}
\vspace{0.5em}
\\
\includegraphics[height=5cm, width=7.cm]{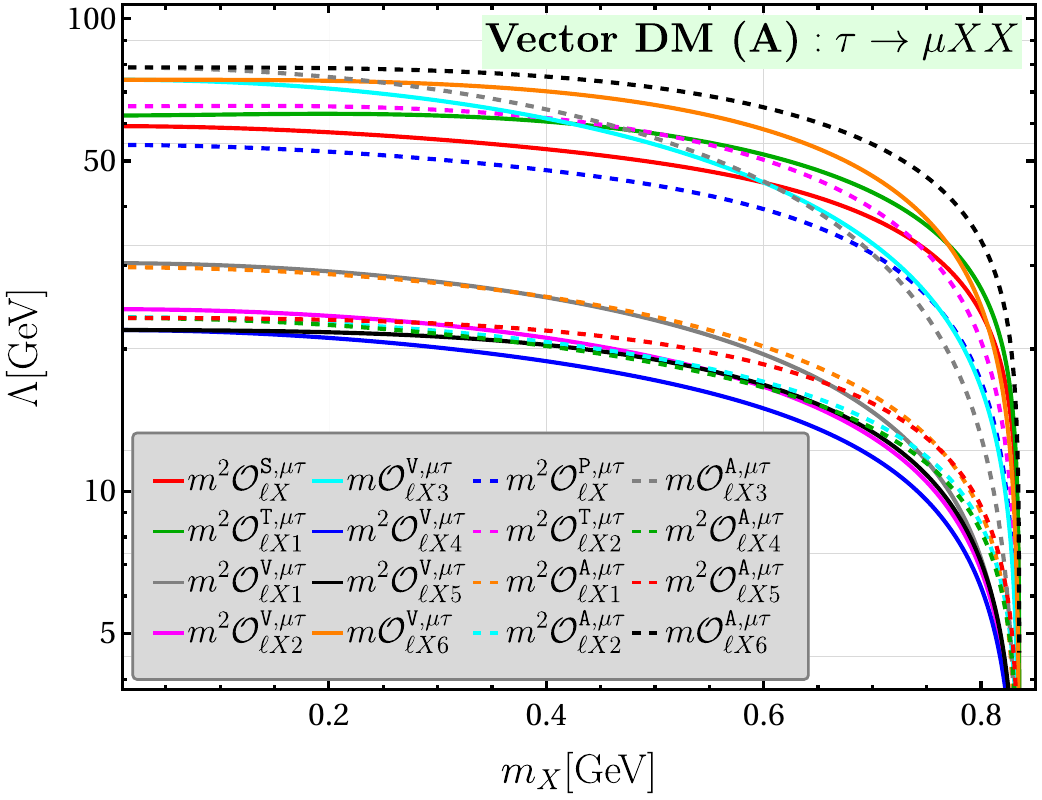}\quad
\includegraphics[height=5cm, width=7.cm]{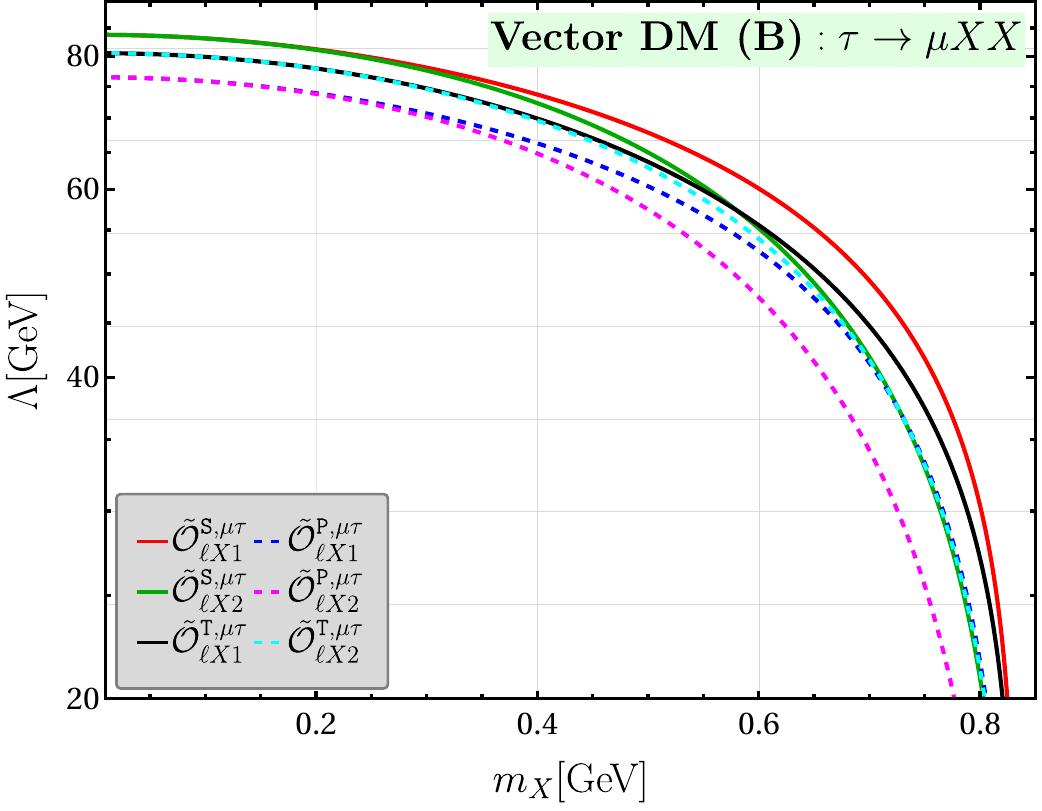}
\caption{Same as \cref{fig:const.mu2e} but for the process $\tau \to \mu +{\tt DM+DM}$.}
\label{fig:const.tau2mu}
\end{figure}

The constraints in the $\Lambda-m$ plane for all the DM scenarios from $\tau \to \mu +{\tt DM+DM}$ decay are displayed in \cref{fig:const.tau2mu}. In this case, we notice a marginal decrement on the constraints compared to those from $\tau \to e +{\tt DM+DM}$ decay for all DM scenarios. Due to the significant mass of the muon (compared to the electron), the degeneracy in the $\Lambda-m$ plane observed in the other two flavor combinations ($e \mu$ and $e\tau$) is now lifted over a wide range of DM masses.

{\bf Scalar DM case}: 
As shown in the top-left panel of \cref{fig:const.tau2mu}, due to less mass splitting between the charged $\tau$ and $\mu$,  we see a small splitting between the variations in the $\Lambda-m_\phi$ plane for the operators $\mathcal{O}^{{\tt S}, \mu \tau}_{\ell \phi}$ and $\mathcal{O}^{{\tt P}, \mu \tau}_{\ell \phi}$.  For the operator $\mathcal{O}^{{\tt S}, \mu \tau}_{\ell \phi}$, when the DM mass is below $\Delta m_{\tau \mu}/4$ ($m_\phi \sim 0.01$ GeV), the constraint on $\Lambda$ is approximately 407 TeV. At $\Delta m_{\tau\mu}/4$ ($m_\phi \sim 0.42$ GeV), the constraint is reduced to 219 TeV. On the other hand, for the operator $\mathcal{O}^{{\tt V}, \mu \tau}_{\ell \phi}$, the constraint on $\Lambda$ is approximately 663 (409) GeV at $m_\phi \sim 0.01~(0.42)$ GeV. A slightly stronger constraint applies to the operator $\mathcal{O}^{{\tt A}, \mu \tau}_{\ell \phi}$ compared to the former.

{\bf Fermion DM case}:
From the top-right panel of \cref{fig:const.tau2mu}, the constraints on $\Lambda$ are marginally reduced by a factor of approximately 1.07 for all operators compared to the case of $\tau \to e \chi\chi$ decay. A significant splitting in the $\Lambda-m_\chi$ variation is identified between the operators $\mathcal{O}^{{\tt S}, \mu \tau}_{\ell \chi 1(2)}$ and $\mathcal{O}^{{\tt P}, \mu \tau}_{\ell \chi 1(2)}$, as well as between $\mathcal{O}^{{\tt V}, \mu \tau}_{\ell \chi 1(2)}$ and $\mathcal{O}^{{\tt A}, \mu \tau}_{\ell \chi 1(2)}$. However, for the operators $\mathcal{O}^{{\tt T}, \mu \tau}_{\ell \chi 1}$ and $\mathcal{O}^{{\tt T}, \mu \tau}_{\ell \chi 2}$, the variations are found to overlap significantly.  The operator $\mathcal{O}^{{\tt T}, \mu \tau}_{\ell \chi 2}$ ($\mathcal{O}^{{\tt P},\mu \tau}_{\ell \chi 1}$) yields the most (least) stringent constraint.  For a DM mass of $ m_\chi \sim 0.01 $ (0.42) GeV, the constraint on $\Lambda$ is approximately 1.51 (1.21) TeV for the operator $\mathcal{O}^{{\tt T}, \mu \tau}_{\ell \chi 1}$. Similarly, for the operator $\mathcal{O}^{{\tt P},\mu \tau}_{\ell \chi 1}$, the constraint on $\Lambda$ is approximately 717 (612) GeV for the same DM mass. 

{\bf Vector DM case}: 
From the bottom-left panel of \cref{fig:const.tau2mu}, the constraints on $\Lambda$ for all case-A operators are marginally reduced by a factor of approximately 1.08 compared to the constraints on the effective scale associated with the $e\tau$ flavor combination. The operator $\mathcal{O}^{{\tt A}, \mu \tau}_{\ell X 6}$ ($\mathcal{O}^{{\tt V},\mu \tau}_{\ell X 4}$) has the most (least) stringent constraint.  For a DM mass of $ m_X \sim 0.01 $ (0.42) GeV, the constraint on $\Lambda$ is approximately 74 (69) GeV for the operator $\mathcal{O}^{{\tt A}, \mu \tau}_{\ell X 6}$. Similarly, for the operator $\mathcal{O}^{{\tt V}, \mu \tau}_{\ell X 4}$, the constraint on $\Lambda$ is approximately 21 (19) GeV for the same DM mass. 

For the case B, unlike $\mu \to e XX$ and $\tau \to e XX$ decays, non-degenerate variations in the $\Lambda-m_X$ plane are noticed for the  operators $\mathcal{\tilde{O}}^{{\tt S}, \mu \tau}_{\ell X 1}$, $\mathcal{\tilde{O}}^{{\tt S}, \mu \tau}_{\ell X 2}$, and $\mathcal{\tilde{O}}^{{\tt T}, \mu \tau}_{\ell X 1}$ from $\tau \to \mu XX$ decay. Variations of $\Lambda$ as a function of $m_X$ for the operators $\mathcal{\tilde{O}}^{{\tt S}, \mu \tau}_{\ell X 1}$, $\mathcal{\tilde{O}}^{{\tt P}, \mu \tau}_{\ell X 1}$, and $\mathcal{\tilde{O}}^{{\tt T}, \mu \tau}_{\ell X 1}$ tend to overlap with the operators $ \mathcal{\tilde{O}}^{{\tt S}, \mu \tau}_{\ell X 2}$, $\mathcal{\tilde{O}}^{{\tt P}, \mu \tau}_{\ell X 2}$, and $\mathcal{\tilde{O}}^{{\tt T}, \mu \tau}_{\ell X 2}$ up to $m_X \sim 0.30~\rm GeV$ and then minutely split after 0.30 GeV. The operator $\mathcal{\tilde{O}}^{{\tt S}, \mu \tau}_{\ell X 1}$ ($\mathcal{\tilde{O}}^{{\tt P},\mu \tau}_{\ell X 2}$) has the most (least) stringent constraint.   For a DM mass of $ m_X \sim 0.01 $ (0.42) GeV, the constraint on $\Lambda$ is approximately 81(66) GeV for the operator $\mathcal{\tilde{O}}^{{\tt S}, \mu \tau}_{\ell X 1}$. Similarly, for the operator $\mathcal{\tilde{O}}^{{\tt P}, \mu \tau}_{\ell X 2}$, the constraint on $\Lambda$ is approximately 76 (63) GeV for the same DM mass. 

As noted at the end of \cref{sec:qsqrd_dist}, for the vector DM case A, we adopt the parameterization of the WCs according to \cref{eq:norm.couplings} to avoid the divergence at $m_X \to 0$. Without this 
reparameterization, the constraint on the $\Lambda$ would be significantly stronger.
For instance, for the operator $\mathcal{O}^{{\tt V}, e \mu}_{\ell X 6}$ the constraint on $\Lambda$ is stronger by 2 (1) orders of magnitude for $m_X \sim 1~(26)$ MeV from the process $\mu^- \to e^- \nu_e \bar{\nu}_{\mu}$. For the process $\tau \to e/\mu + \text{inv.}$, the constraint on $\Lambda$ associated with the operator $\mathcal{O}^{{\tt V}, e\tau}_{\ell X 6}/\mathcal{O}^{{\tt V}, \mu \tau}_{\ell X 6}$ is approximately enhanced by more than  3 (2) orders of magnitude for $m_X \sim 0.01~(0.44)$ GeV.

\begin{figure}[b]
\centering
\includegraphics[width=0.32\textwidth]{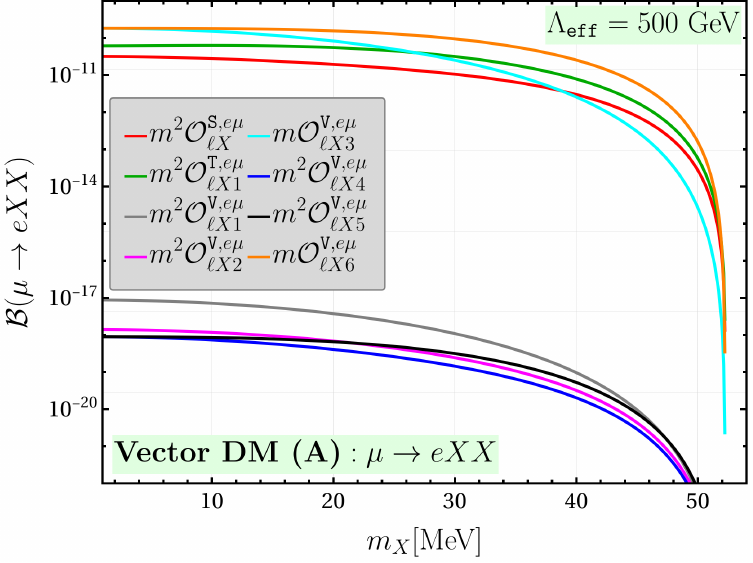}~
\includegraphics[width=0.32\textwidth]{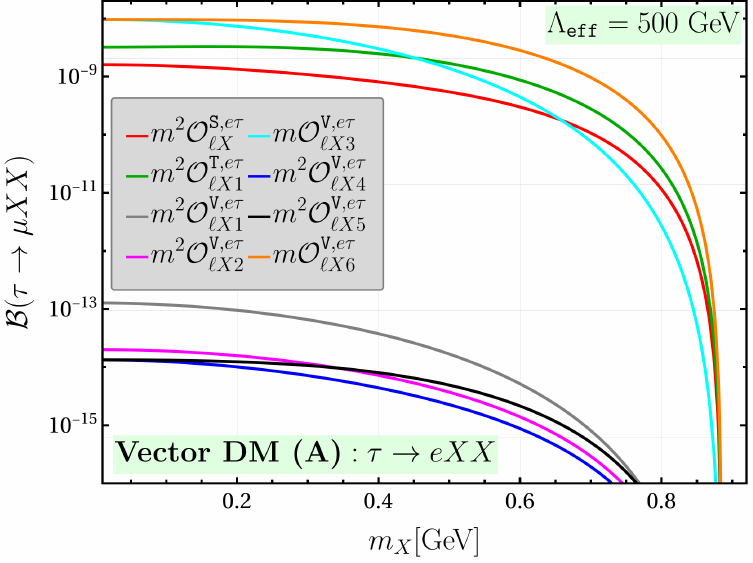}~
\includegraphics[width=0.32\textwidth]{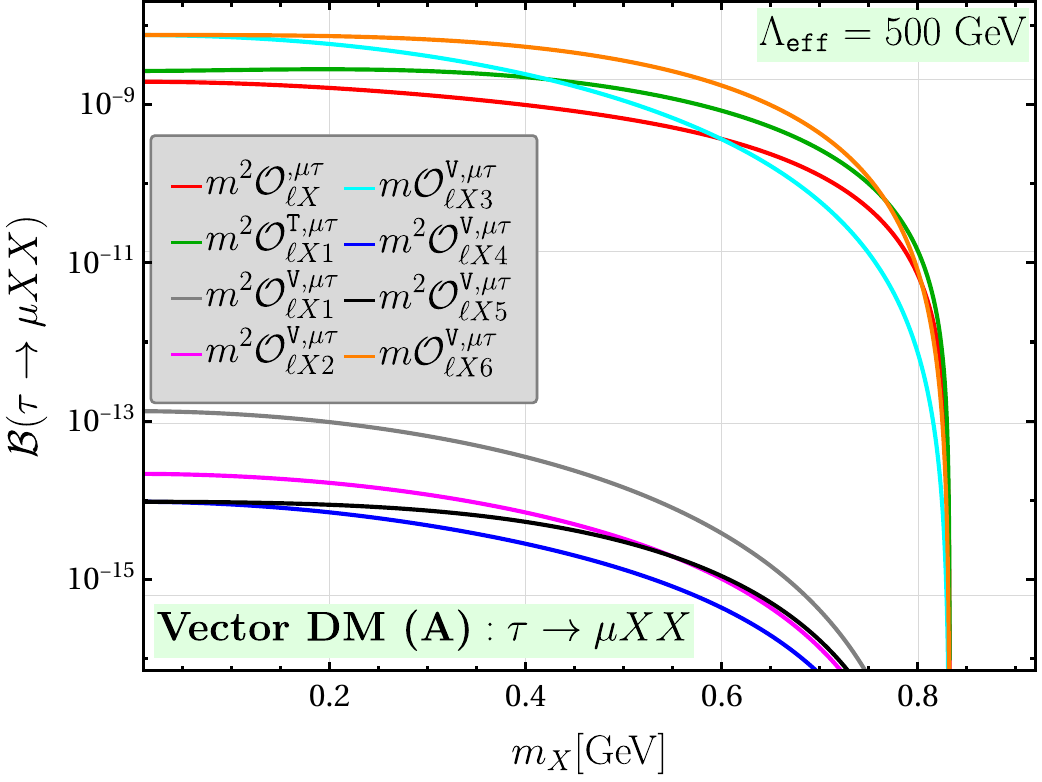}
\caption{Prediction of branching ratios in vector DM case A for the fixed effective scale $\Lambda_{\text{eff}}=500$ GeV. }
\label{fig:br.VDMA}
\end{figure}

{\bf Prediction of branching ratios in vector DM case A}: 
Due to the weak constraints on $\Lambda$ arising from the reparametrization in \cref{eq:norm.couplings}, we instead estimate the branching ratios for a fixed effective scale $\Lambda_{\text{eff}} = 500~\text{GeV}$. \cref{fig:br.VDMA} presents the predicted branching ratios for the three processes ($\mu \to e XX$, $\tau \to e XX$, and $\tau \to \mu XX$) induced by various operators. For simplicity, we consider only the {\tt S/V/T1} types of operators, as the other half of operators with {\tt P/A/T2} structures yield very similar results due to the fact $m_j \ll m_i$. 
Among the operators considered, $\mathcal{O}^{{\tt V}, ji}_{\ell X 6}$ consistently gives the largest contribution to the branching ratios, while $\mathcal{O}^{{\tt V}, ji}_{\ell X 4}$ provides the smallest.
For the process $\mu \to e XX$, the branching ratio is approximately $1.86~(1.31) \times 10^{-10}$ for a DM mass of $m_X \sim 1~(26)~\text{MeV}$  when considering the operator $\mathcal{O}^{{\tt V}, e\mu}_{\ell X 6}$. In contrast, the operator $\mathcal{O}^{{\tt V}, e\mu}_{\ell X 4}$ yields a much smaller branching ratio of around $9.05~(2.61) \times 10^{-19}$ for the same mass points. In the case of $\tau \to e XX$, for $m_X \sim 0.01~(0.44)~\text{GeV}$, the branching ratio is estimated to be $9.67~(6.46) \times 10^{-9}$ with the operator $\mathcal{O}^{{\tt V}, e\tau}_{\ell X 6}$, and $1.34~(0.35) \times 10^{-14}$ for $\mathcal{O}^{{\tt V}, e\tau}_{\ell X 4}$. Finally, for the process $\tau \to \mu XX$, the branching ratio is approximately $7.56~(5.04) \times 10^{-9}$ for $m_X \sim 0.01~(0.42)~\text{GeV}$ with the operator $\mathcal{O}^{{\tt V}, \mu\tau}_{\ell X 6}$, while it reduces to $9.56~(2.47) \times 10^{-15}$ for the operator $\mathcal{O}^{{\tt V}, \mu\tau}_{\ell X 4}$.

\section{Muon four-body radiative decay $\mu\to e+\tt{DM+DM}+\gamma$}
\label{sec:4Bdecay}
Muon radiative decay is one of the important channels to look for light NP. The branching ratio of the process $\mu^- \to e^- \nu_\mu \bar\nu_e \gamma$ has been measured precisely at the MEG experiment \cite{MEG:2013mmu}. Since the neutrinos are treated as missing energy in experimental detection, the presence of any invisible particles other than a neutrino pair is also possible. In this section, we investigate the sensitivity of probing LFV DM interactions involving the $e\mu$ flavor combination through the four-body process $\mu\to e+\tt{DM+DM}+\gamma$. We will analyze the missing energy distribution for the process and take the uncertainty in the measured branching ratio under the same experimental conditions to establish independent new limits for the effective operators.  Specifically, we require ${\cal B}(e+{\tt DM+DM}+\gamma)\lesssim 6.03\times 10^{-8}$ with the conditions that the electron and photon energies satisfy the MEG restrictions, $E_e>45~\rm MeV$ and $E_\gamma >40~\rm MeV$.

\subsection{Event distributions}

\begin{figure}[b]
\centering
\includegraphics[width=0.48\textwidth]{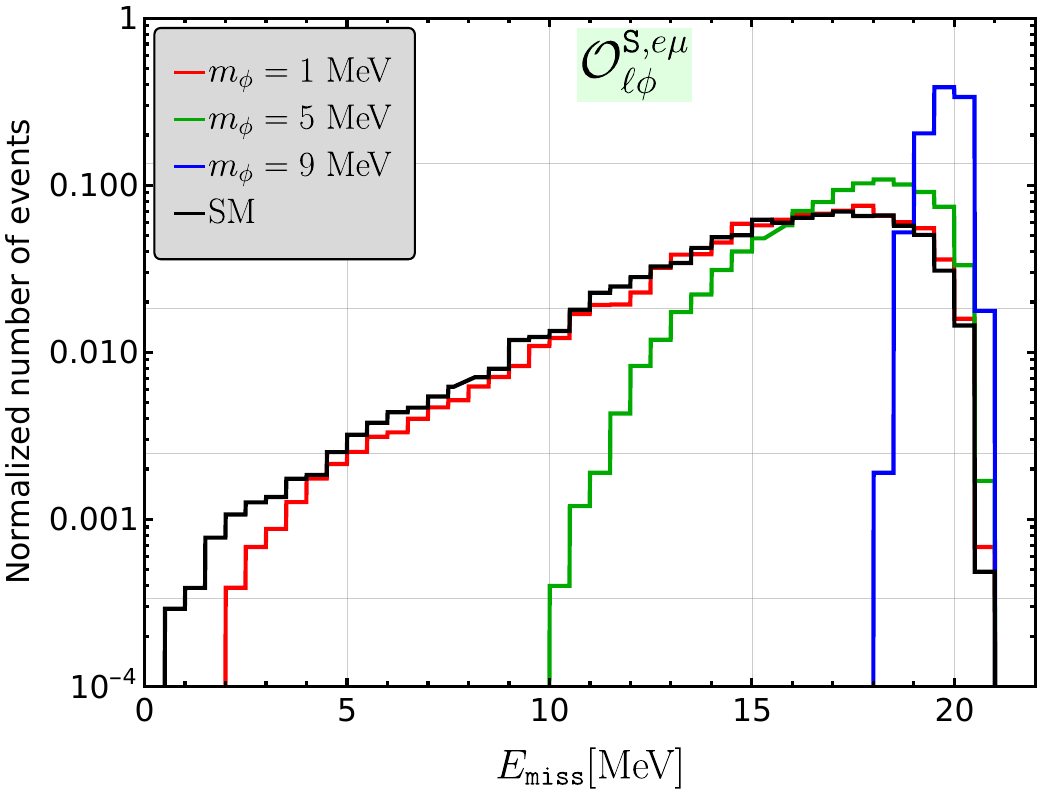}~~
\includegraphics[width=0.48\textwidth]{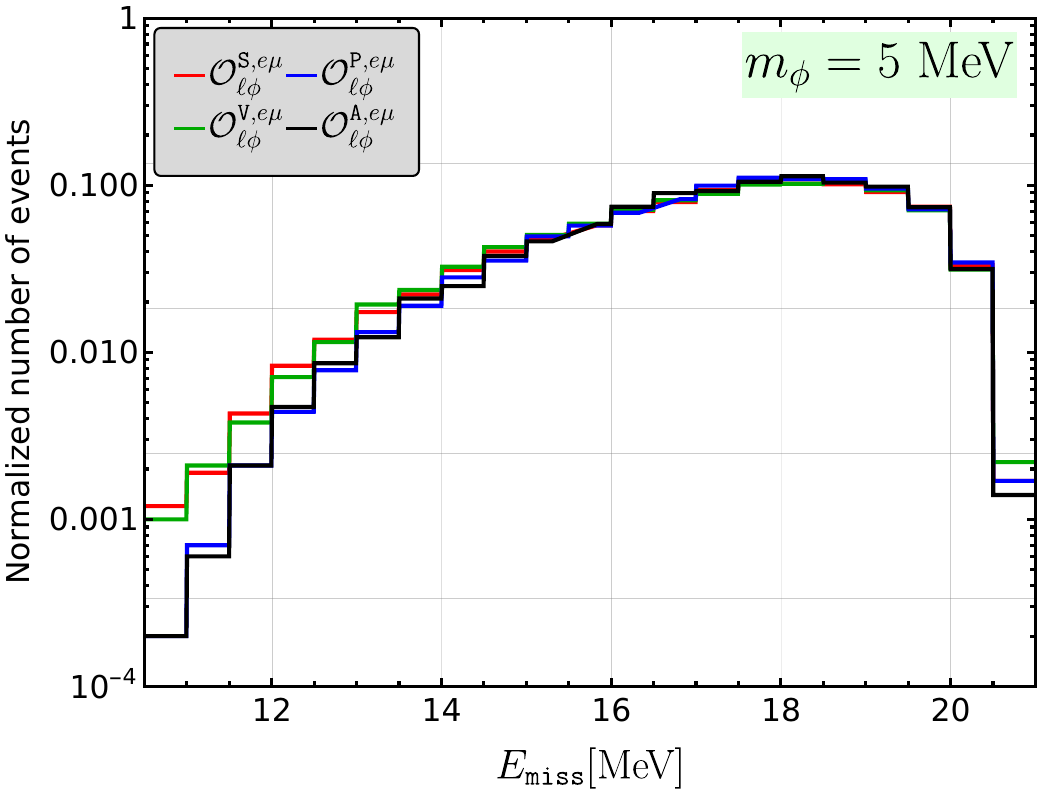}
\caption{
Missing energy distribution for the process $\mu \to e \phi \phi \gamma$ in the scalar DM scenario across various benchmark cases.}
\label{fig:emiss}
\end{figure}

The missing energy ($E_{\text{miss}}$) is a crucial kinematic variable for investigating scenarios involving invisible new particles across a wide range of experimental setups. The definition of $E_{\text{miss}}$ in the case of radiative four-body muon decay is given by
\begin{align}
E_{\text{miss}}=m_{\mu}-E_e -E_{\gamma}.
\label{eq:e.miss}
\end{align}
In the left plot of \cref{fig:emiss}, we present the normalized event distribution for $E_{\text{miss}}$ for several DM masses, focusing on the scalar DM case with the operator $\mathcal{O}^{{\tt S}, e \mu}_{\ell \phi}$. For comparison, we also include the SM distribution for the process $\mu^- \to e^- \nu_\mu \bar\nu_e \gamma$. In the plots, we have applied the aforementioned experimental selection cuts on electron and photon energies, which restrict the endpoint of the distribution to approximately 20 MeV. \cref{eq:e.miss} reveals that the $E_{\text{miss}}$-distribution for a DM mass $m$ begins at $2m$ and extends to roughly 20 MeV, as illustrated in \cref{fig:emiss}. Hence, similar to the $q^2$-distribution for the three-body decay modes, the initial point of $E_{\text{miss}}$ could be used to determine the mass of the DM in the future similar experiments.  The distribution further demonstrates that increasing DM mass enhances the separation of the NP signal from the SM background. However, from the right plot of \cref{fig:emiss}, we find that this distribution does not differentiate between operators with distinct Lorentz structures very well, when the missing energy is cut to a larger value. 

\begin{figure}[t]
\centering
\includegraphics[width=0.48\textwidth]{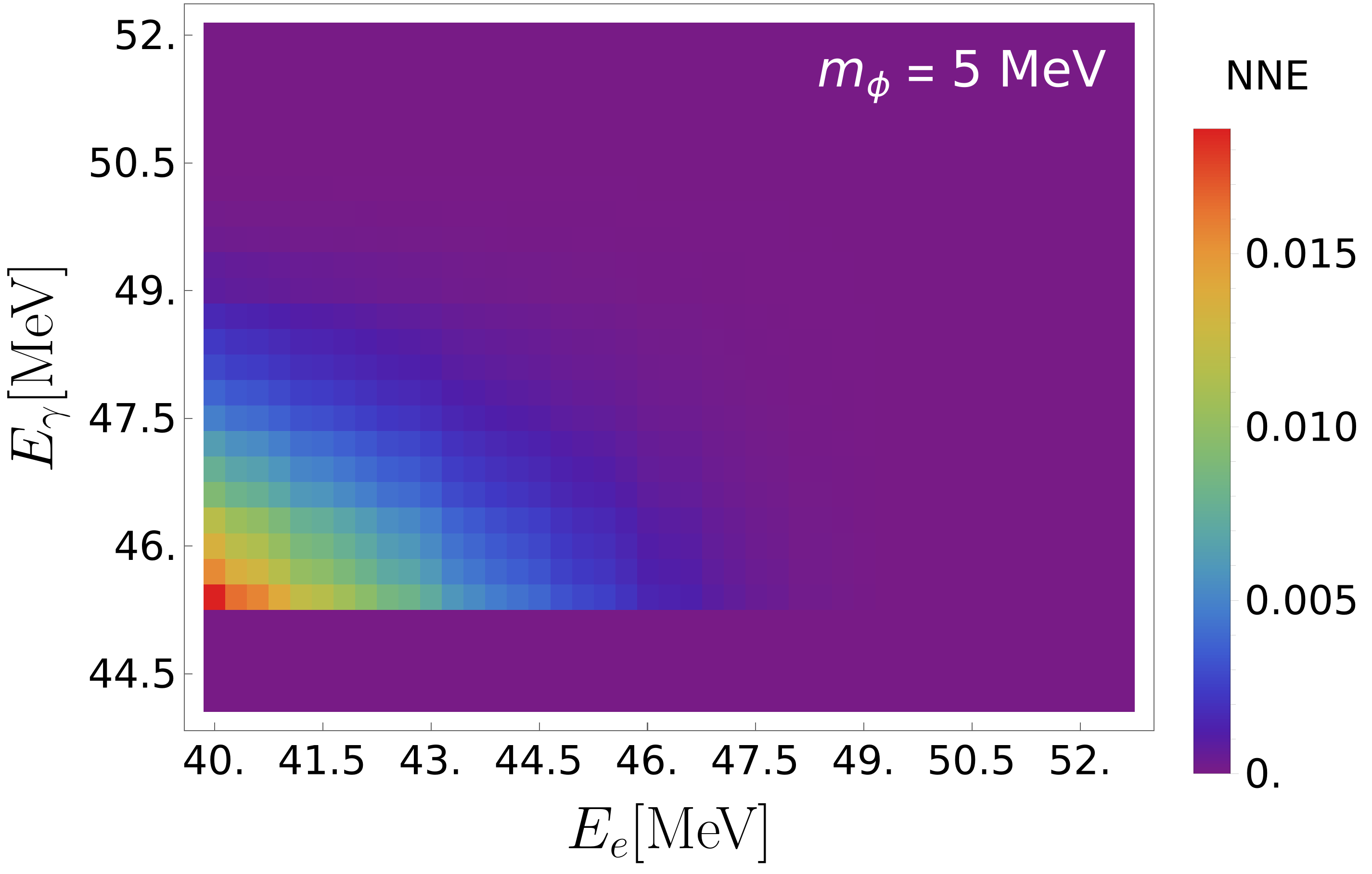}~~
\includegraphics[width=0.48\textwidth]{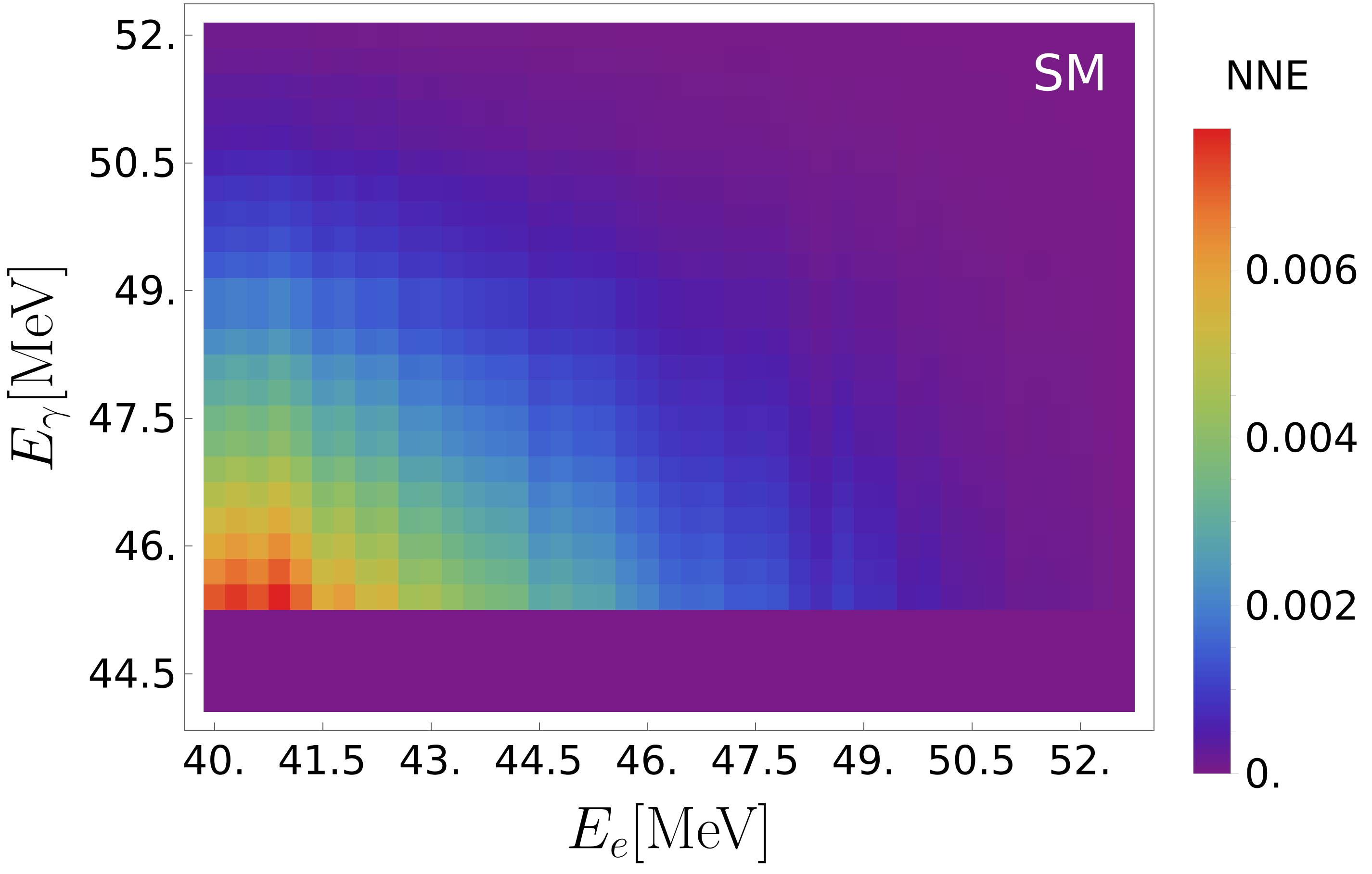}
\caption{Two-dimensional distribution of the normalized number of events (NNE) in the $E_e{\rm-}E_{\gamma}$ plane for the decay $\mu \to e \phi \phi \gamma$ due to the operator $\mathcal{O}^{{\tt S}, e \mu}_{\ell \phi}$ in the scalar DM case (left) and for the SM decay (right).}
\label{fig:emiss_2d}
\end{figure}

The normalized two-dimensional event distribution in the $E_e{\rm-}E_{\gamma}$ plane is shown in the left plot of \cref{fig:emiss_2d} for the operator $\mathcal{O}^{{\tt S}, e \mu}_{\ell \phi}$ 
with $m_{\phi}=5$ MeV. 
For comparison, we also present the event distribution of the corresponding SM process on the right panel.
Events with highly energetic $E_{\gamma}$ and $E_e$ are less favored, as most of the energy is carried away by the DM (neutrino) pair for the DSEFT (SM) scenario. Within the SM, the event distribution extends up to approximately 52.5 MeV for the $E_e$ and 52 MeV for the $E_{\gamma}$. In contrast, within the DSEFT framework, the distribution is less extended---reaching only up to about 49.5 MeV for $E_e$ and 50 MeV for $E_{\gamma}$---due to the lower cuts on $E_{e,\gamma}$ and a larger DM mass $m_\phi\gg m_\nu$.

\subsection{Constraints from the four-body radiative decay}

\begin{figure}[t]
\centering
\includegraphics[width=7.cm]{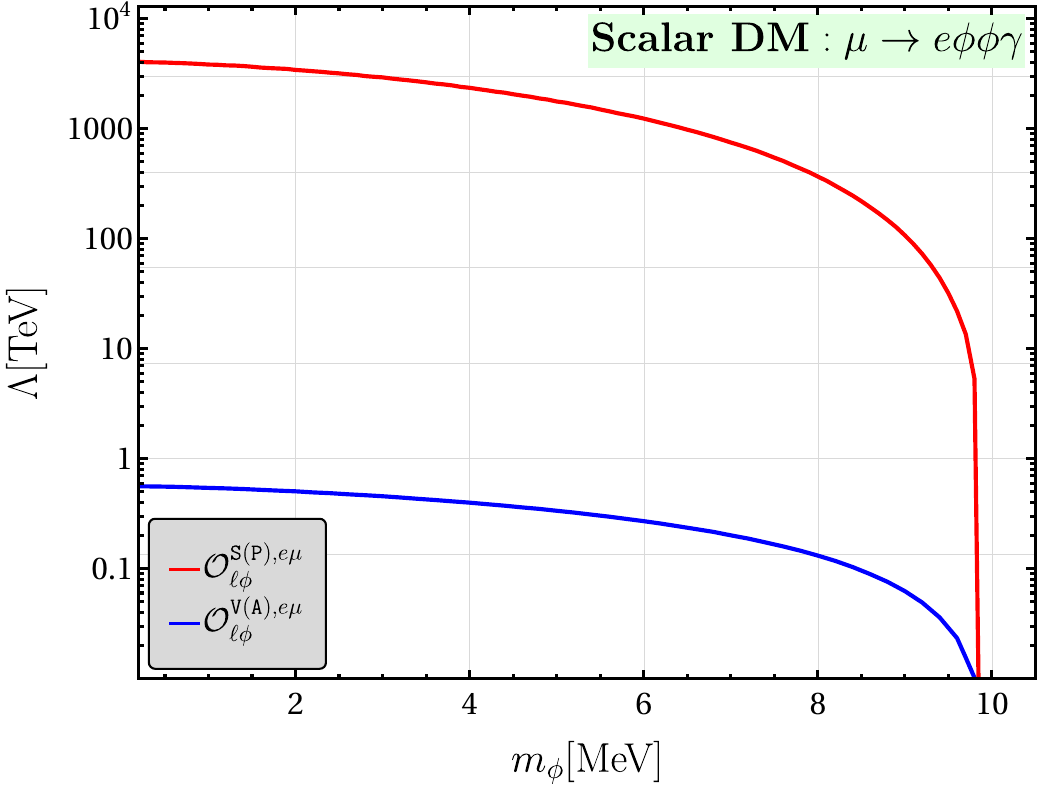}\quad
\includegraphics[width=7.cm]{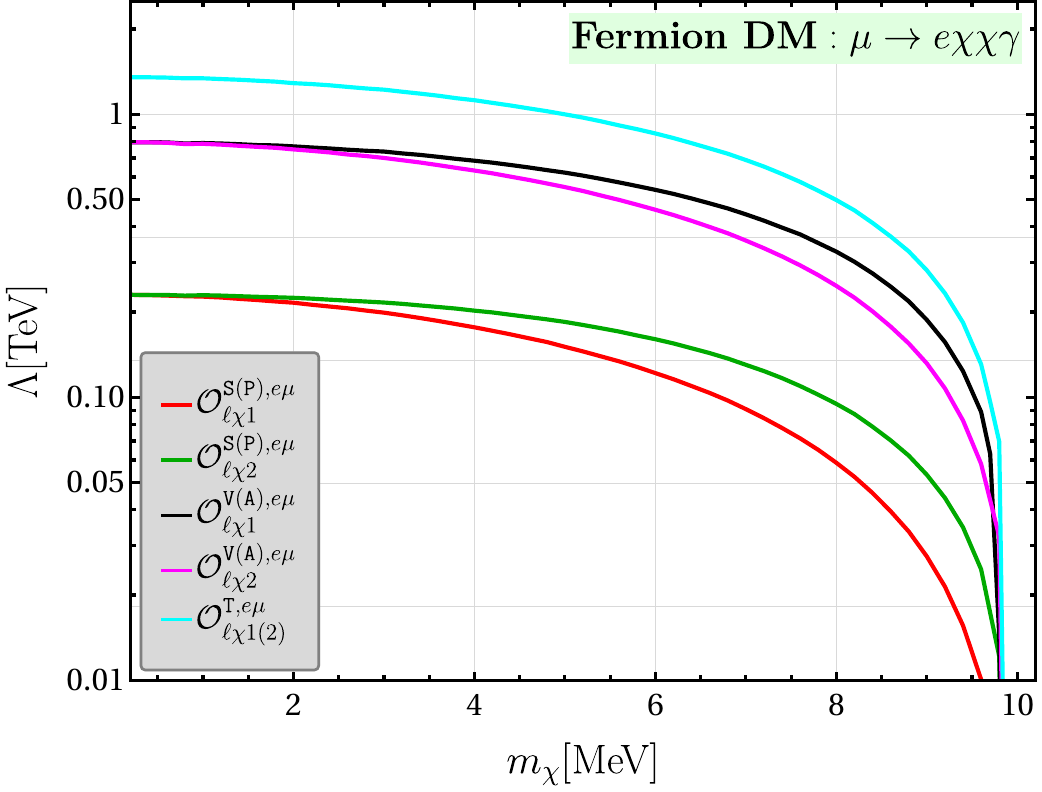}
\vspace{0.5em}
\\
\includegraphics[width=7.cm]{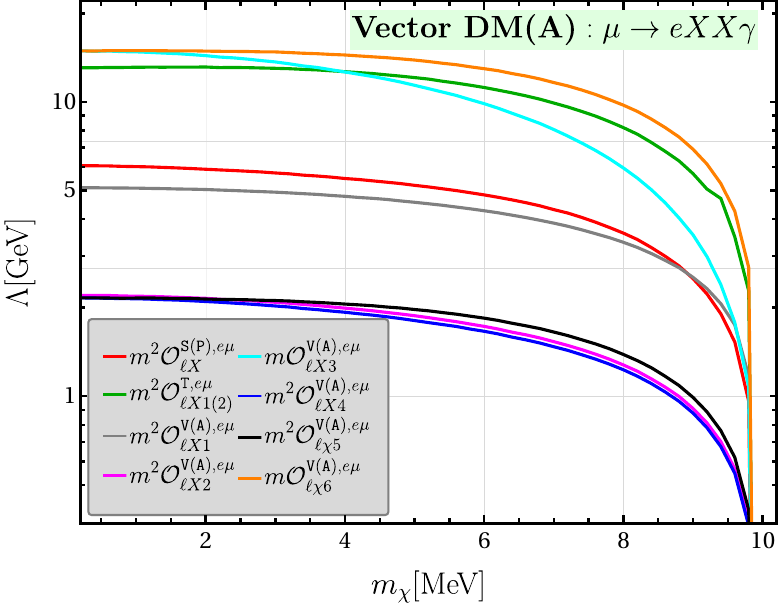}\quad
\includegraphics[width=7.cm]{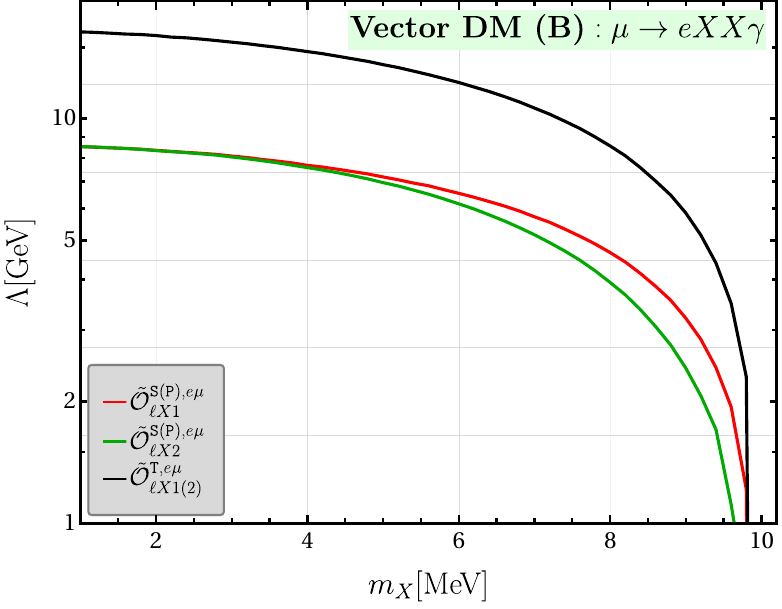}
\caption{Same as \cref{fig:const.mu2e} but for the radiative process $\mu \to e +{\tt DM+DM}+ \gamma$.}
\label{fig:4body.rad}
\end{figure}

To calculate the four-body decay width based on the DSEFT operators in \cref{tab:operators}, we implement these interactions in {\tt Feynrules} \cite{Alloul:2013bka} to generate UFO files, which are then used in {\tt Madgraph} \cite{Alwall:2011uj} to evaluate the constraints in the $\Lambda-m$ plane. After employing the selection cuts on the electron ($E_e > 45$ MeV) and photon ($E_{\gamma}>40$ MeV) following the MEG experiment, the constraints in the $\Lambda-m$ plane for all operators in all DM scenarios are shown in \cref{fig:4body.rad}. After these selection cuts, it is understood from the energy conservation that the DM mass can only be probed up to $10~\rm MeV$. In the scalar DM scenario, for the operators $\mathcal{O}^{{\tt P}, e \mu}_{\ell \phi}$ and $\mathcal{O}^{{\tt S}, e \mu}_{\ell \phi}$, this radiative decay provides similar constraints in the $\Lambda-m_\phi$ plane for DM masses up to 2 MeV, when compared to the three-body decay $\mu \to e \phi\phi$. Beyond that mass range, the decay $\mu \to e \phi\phi$ provides a better constraint on $\Lambda$. For the remaining operators in other DM scenarios, the decay $\mu \to e+{\tt DM+DM}$ provides better constraints for the allowed mass range. In the scalar DM scenario, the operators $\mathcal{O}^{{\tt S/P}, e \mu}_{\ell \phi}$ ($\mathcal{O}^{{\tt V/A}, e \mu}_{\ell \phi}$) have the most (least) stringent constraint.  For a DM mass of $ m_\phi \sim 0.1 $ (5) MeV, the constraint on $\Lambda$ is approximately 4050 (1770) TeV for the operators $\mathcal{O}^{{\tt S/P}, e \mu}_{\ell \phi}$. Similarly, for the operators $\mathcal{O}^{{\tt V/A}, e \mu}_{\ell \phi}$, the constraint on $\Lambda$ is approximately 560 (335) GeV for the same DM mass. For the fermion DM scenario,  the operators $\mathcal{O}^{{\tt T}, e \mu}_{\ell \chi 1/2}$ ($\mathcal{O}^{{\tt S/P}, e \mu}_{\ell \chi 1}$) have the most (least) stringent constraint.  For a DM mass of $ m_\chi \sim 0.1 $ (5) MeV, the constraint on $\Lambda$ is approximately 1.35 (1) TeV for the operators $\mathcal{O}^{{\tt T}, e \mu}_{\ell \chi 1/2}$. Similarly, for the operators $\mathcal{O}^{{\tt S/P}, e \mu}_{\ell \chi 1}$, the constraint on $\Lambda$ is approximately 230 (150) GeV for the same DM mass. 
Unfortunately, due to the parameterization defined in \cref{eq:norm.couplings}, the constraints are significantly weaker for the operators $\calO_{\ell X2,4,5}^{\tt V(A),e\mu}$ in vector DM case A than in other scenarios and therefore become meaningless.
In the vector DM case B, the relatively weaker bounds arise from the higher dimensionality of the operators.

\section{Implications on muonium invisible decay}
\label{sec:Mmudecay}

The muonium is a QED bound state consisting of a positively charged muon and an electron, $|M_{\mu}\rangle=|\mu^+e^-\rangle$. There are two muonium states based on the spin configurations: the spin singlet is termed as para-muonium ($M_\mu^P$) while the triplet is named as ortho-muonium ($M_\mu^O$). Search for invisible decays of muonium is driven by compelling motivations rooted in both the SM and potential NP. The SM decay $M_\mu(\mu^+ e^-) \to \nu_e \bar{\nu}_{\mu}$ is predicted to occur at a rate accessible to current experimental techniques, making its observation a fascinating test of the SM.  Any deviation from the SM rate could indicate the presence of low-mass DM, providing a unique opportunity to explore BSM physics and uncover potential signatures of DM.

Since muon is unstable, $M_{\mu}$ is also unstable. Within the SM, the main decay channel of $M_{\mu}$ coincides with that of the muon, $\mu^+ \to e^+ \nu_{e} \bar{\nu}_{\mu}$. Therefore, the lifetime of the muonium is approximately taken to be $\tau_{M_{\mu}}\approx 1/\Gamma (\mu^+ \to e^+ \nu_{e} \bar{\nu}_{\mu}$). The SM predicts the direct annihilation of the ortho-muonium into a neutrino-antineutrino pair with a very small decay rate. The corresponding branching ratio has been calculated to be \cite{Li:1988xb}
\begin{align}
{\cal B}(M_\mu^O \to \nu_e \bar{\nu}_{\mu}) \equiv 
\frac{\Gamma(M_\mu^O \to \nu_e \bar{\nu}_{\mu})}{\Gamma(\mu^+ \to e^+ \nu_e \bar{\nu}_{\nu})}
=48 \pi \alpha^3 \left(\frac{m_e}{m_{\mu}}\right)^3
\simeq 6.6 \times 10^{-12},
\label{eq:br.Mu.SM}
\end{align}
where $\alpha$ is the fine structure constant. 
Based on the precision measurements of the positive muon lifetime from the MuLan collaboration \cite{MuLan:2010shf,MuLan:2012sih}, an upper bound on the branching ratio of invisible decay of ortho-muonium  has been evaluated as \cite{Gninenko:2012nt}
\begin{equation}
{\cal B}(M_\mu^O  \to \text{inv.}) < 5.7 \times 10^{-6}\quad 
@\, 90\%\,{\rm C.L.}.
\end{equation}

In the case of muonium invisible decay, non-vanishing matrix elements due to axial-vector, vector, and tensor leptonic currents are parametrized as 
\begin{align}
&\langle 0|\bar{\mu}\gamma^{\alpha} e|M_{\mu}^O \rangle 
= i f_V M_{M}\epsilon_M^{\alpha}, \qquad 
\langle 0|\bar{\mu}\sigma^{\alpha \beta} e|M_{\mu}^O \rangle 
= i f_T (\epsilon_M^{\alpha} p^{\beta}-\epsilon_M^{\beta} p^{\alpha}), \\
&\langle 0|\bar{\mu} \gamma_5 e|M_{\mu}^P \rangle
= -i f_P M_M, \qquad 
\langle 0|\bar{\mu}\gamma^{\alpha} \gamma_5 e|M_{\mu}^P \rangle 
= i f_P p^{\alpha}, 
\label{eq:decay.cons}
\end{align}
where $p$ is the four-momentum of the initial muonium state whose mass is denoted by $M_M \approx m_\mu$, while $\epsilon_M$ is the polarization vector of the ortho-muonium state. Except for the pseudoscalar current, other matrix elements can be found in \cite{Petrov:2022wau}.
$f_P$, $f_V$, and $f_T$ are the muonium decay constants corresponding to the axial-vector, vector, and tensor leptonic currents, respectively. In the non-relativistic limit, $f_P=f_V=f_T \equiv f_M$. The decay constant $f_M$ can be written in terms of the bound-state wavefunction by employing the QED version of the Van Royen-Weisskopf formula,
\begin{equation}
f_M^2=4\frac{|\phi(0)|^2}{M_M}.
\end{equation}
The absolute value of the wavefunction at the origin, $\phi(0)$, can be expressed as 
\begin{equation}
|\phi(0)|^2=\frac{(m_{\text{red}}\alpha)^3}{\pi}, \quad  m_{\text{red}}=\frac{m_{e} m_{\mu}}{m_e + m_{\mu}}.
\end{equation}

\begin{figure}[b]
\centering
\includegraphics[width=0.47\textwidth]{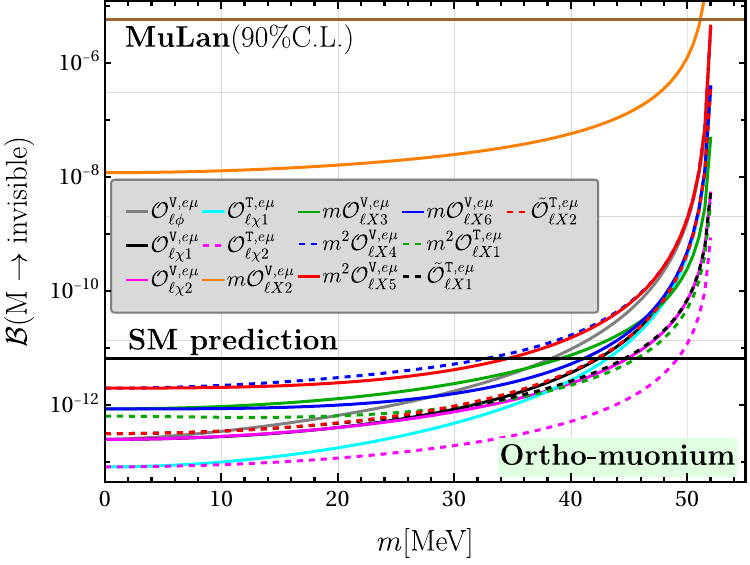}~~
\includegraphics[width=0.47\textwidth]{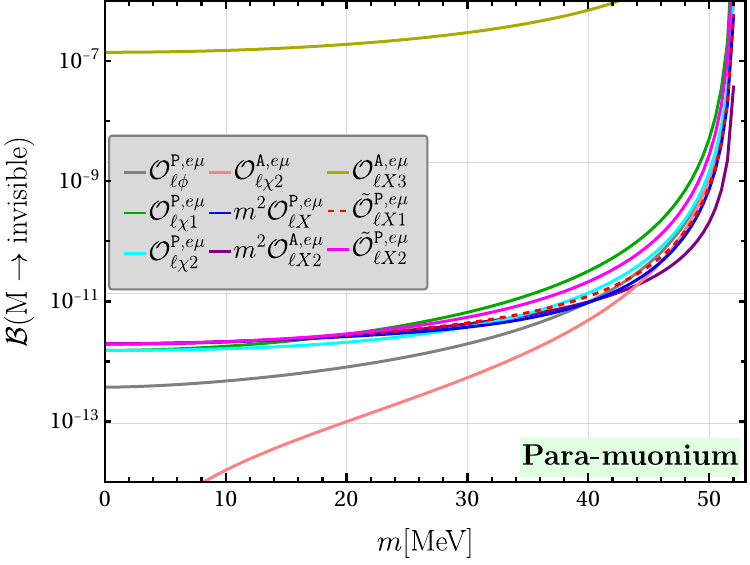}
\caption{Limit on branching ratio of muonium invisible decay from the insertion of various operators.}
\label{fig:br.muonimum}
\end{figure}

The decay width for the ortho-muonium invisible decay in each DM scenario is expressed as \footnote{We do not consider the operators $\mathcal{O}^{{\tt V}, e \mu}_{\ell X 1}$ and $\mathcal{O}^{{\tt A}, e \mu}_{\ell X 1}$ as there is no available decay constant for these lepton currents.}
\begin{subequations}
\label{eq:OMu_inv}
\begin{align}
 \Gamma_{M_\mu^O \to \phi\phi} & = 
 \frac{f_M^2 M_M^3 }{12 \pi} (1-4 z_\phi )^{3 \over 2} |C^{{\tt V}, e\mu}_{\ell \phi}|^2, 
\\
 \Gamma_{M_\mu^O \to \chi\chi} &= 
 \frac{f_M^2 M_M^3 }{12 \pi} (1-4 z_\chi)^{1\over2} \Big[ (1+2 z_\chi ) |C^{{\tt V}, e\mu}_{\ell \chi 1}|^2+ (1 -4 z_\chi )|C^{{\tt V}, e\mu}_{\ell \chi 2}|^2
 \nonumber\\
 & +2 (1 +8 z_\chi )|C^{{\tt T}, e\mu}_{\ell \chi 1}|^2
 +2 (1-4 z_\chi) |C^{{\tt T}, e\mu}_{\ell \chi 2}|^2\Big]+\cdots, 
\\
 \Gamma^A_{M_\mu^O \to X X} & = \frac{f_M^2 M_M^3 }{192 \pi} z_X^{-2} (1-4 z_X)^{1\over2}  \Big[ (1-16 z_X^2) M_M^{-2} |C^{{\tt T}, e\mu}_{\ell X 1}|^2
+ 4 z_X (1-4 z_X)|C^{{\tt V}, e\mu}_{\ell X 2}|^2
 \nonumber\\
 & +4 z_X (1 -4 z_X )^2|C^{{\tt V}, e\mu}_{\ell X 3}|^2
 +(1-4 z_X )(1-4 z_X +12 z_X^2 )|C^{{\tt V}, e\mu}_{\ell X 4}|^2
 \nonumber\\
 &  +(1-16 z_X^2 )|C^{{\tt V}, e\mu}_{\ell X 5}|^2
 +z_X (1+2 z_X)|C^{{\tt V}, e\mu}_{\ell X 6}|^2\Big]+\cdots,
\\
 \Gamma^B_{M_\mu^O \to X X} & = \frac{f_M^2 M_M^5 }{96 \pi}
 (1-4 z_X)^{1\over2}
 \left[(1-2 z_X -8 z_X^2)|\tilde{C}^{{\tt T}, e\mu}_{\ell X 1}|^2 + (1-2 z_X+4 z_X^2)|\tilde{C}^{{\tt T}, e\mu}_{\ell X 2}|^2\right],
\end{align}
\end{subequations}
where $z_i \equiv m_i^2/M_M^2$ with $i=\phi,~\chi,~X$, and the dots denote the interference terms. For the para-muonium, the non-vanishing ones are
\begin{subequations}
\label{eq:PMu_inv}
\begin{align}
\Gamma_{M_\mu^P \to \phi\phi} & = \frac{f_M^2 M_M}{16 \pi}(1-4 z_\phi)^{1\over2}|C^{{\tt P},e\mu}_{\ell \phi}|^2,
\\
 \Gamma_{M_\mu^P \to \chi\chi} & = 
 \frac{f_M^2 M_M^3 }{8 \pi} (1-4 z_\chi)^{1\over2} \left[ (1-4z_{\chi}) |C^{{\tt P},e\mu}_{\ell \chi1}|^2 + |C^{{\tt P},e\mu}_{\ell \chi 2}|^2 +  4 z_\chi |C^{{\tt A},e\mu}_{\ell \chi2}|^2 \right],
\\
\Gamma^A_{M_\mu^P \to X X} & = \frac{f_M^2 M_M^3}{64 \pi} z_X^{-2} (1-4 z_X)^{1\over2}  \Big[ \frac{(1-4z_X+12z_X^2)}{M_M^2} |C^{{\tt p}, e\mu}_{\ell X }|^2 + (1-4 z_X)^2 |C^{{\tt A}, e\mu}_{\ell X 2}|^2\\\nonumber
&+8z_X^2 (1-4 z_X) |C^{{\tt A}, e\mu}_{\ell X 3}|^2\Big],
\\
\Gamma^B_{M_\mu^P \to X X} & = 
\frac{f_M^2 M_M^5}{16 \pi} (1-4 z_X)^{1\over2} \left[(1-4z_X+6z_X^2)|\tilde{C}^{{\tt P}, e\mu}_{\ell X 1}|^2 + (1-4z_X)|\tilde{C}^{{\tt P}, e\mu}_{\ell X 2}|^2\right].
\end{align}
\end{subequations}
Using \cref{eq:OMu_inv,eq:PMu_inv}, the upper bounds on the invisible decay branching ratios of ortho- and para-muonium are shown in the left and right panels of \cref{fig:br.muonimum}, respectively. These branching ratios are calculated based on constraints from $\mu \to e +{\tt DM+DM}$ decay in previous sections, under the assumption of a single operator being inserted each time. 
From \cref{eq:decay.cons}, it is evident that vector and tensor leptonic current operators contribute to ortho-muonium decay, while pseudoscalar and axial-vector leptonic current operators are relevant for para-muonium decay.
The remaining operators with non-vanishing leptonic matrix elements that 
do not contribute to the ortho/para-muonium invisible decay are due to kinematics governed by the DM Lorentz structures.
In addition, when restricting to the real/Majorana DM scenario, only the fermionic DM operator $\calO_{\ell\chi2}^{{\tt V},e\mu}$ and the vector DM operators $\calO_{\ell X2,3}^{{\tt V},e\mu}$ survive for ortho-muonium; 
whereas, all operators in \cref{eq:PMu_inv} still survive for para-muonium case. 

For ortho-muonium invisible decay, the mass range below the black solid line for each relevant operator in \cref{fig:br.muonimum} predicts a branching ratio smaller than the SM value, making it challenging to observe in future experiments. The operator $\mathcal{O}^{{\tt V},e \mu}_{\ell X 2}$ in the vector DM case A provides the largest contribution to the branching ratio, while the fermionic DM operator $\mathcal{O}^{{\tt T},e \mu}_{\ell \chi 2}$ offers the smallest contribution due to the stringent constraint from $\mu\to e +{\tt DM+DM}$ decay. For para-muonium invisible decay, the operator $\mathcal{O}^{{\tt A},e \mu}_{\ell X 3}$ ($\mathcal{O}^{{\tt A},e \mu}_{\ell \chi 2}$) provides the largest (smallest) contribution.
From kinematic constraints, the upper bound on $m$ is approximately 52.5 MeV. An intriguing aspect is that, due to angular momentum conservation, para-muonium cannot undergo any two-body invisible decay within the SM, and its four-body invisible decay is highly suppressed. Consequently, any observation of invisible decay from para-muonium would serve as a compelling {\it smoking gun} signature of these NP interactions.

\section{Summary}
\label{sec:summary}

In this paper, we have systematically investigated potential lepton flavor violating (LFV) interactions between charged leptons and dark matter (DM) particles. As a preliminary investigation in this direction, we focused on a light DM candidate and considered the LFV muon and tau decays into a lighter lepton and a pair of invisible DM particles in the final state to probe such flavored DM scenarios. This analysis was conducted within the low energy DSEFT framework. We have collected all leading-order DSEFT operators that consist of a flavor-changing charged lepton pair and a pair of DM fields, encompassing the three well-studied scenarios of scalar, fermion, and vector DM. Notably, the case involving two DM particles in the charged LFV decay processes has not been explored in previous studies. We have examined this case by analyzing the invariant mass distribution for the decays and establishing constraints on the associated effective scale for each operator based on current experimental limits from similar processes for all DM scenarios. Furthermore, we have extended our analysis to the invisible decay of the muonium atom within the same DSEFT framework, establishing new upper bounds on the branching ratios for muonium invisible decays by utilizing the limits obtained from charged LFV decays. This investigation holds significant importance for exploring charged LFV processes in future low energy experiments. Our study is particularly significant in scenarios where new symmetries prevent the production of single DM particles.
It should be emphasized that all results obtained in this work are also applicable to scenarios involving a pair of light dark particles beyond the DM framework.

We found that the differential $q^2$-distribution of three-body decay $\ell_i \to \ell_j+{\tt DM+DM}$ is pivotal for distinguishing between different DSEFT operators and for understanding the Lorentz structures of the leptonic currents associated with these operators. 
Among the three processes studied, the decay $\tau \to \mu + {\tt DM + DM}$ emerges as the most effective in differentiating the chiral Lorentz structures of the leptonic currents due to the sizable muon mass. 
Additionally, the endpoint of $q^2$-distribution can be used to determine the value of the DM mass. Regarding the constraints in the $\Lambda-m$ plane derived from the upper bounds of flavor-violating muon and tau decays, we found that the process $\mu \to e+{\tt DM+DM}$ provides the most stringent constraints on the effective scale associated with operators $\calO^{{\tt S(P)},e\mu}_{\ell \phi}$. 
For tau-related operators, both tau decay processes impose comparable constraints on the effective scales of operators sharing the same Lorentz structure. It is worth noting that the constraint depends not only on the DM mass but also on the flavor structure of the WC.
Furthermore, we have also considered the constraints from the four-body muon radiative decay in the $e \mu$ flavor combination. Below $\Delta m_{\mu e}/4$, we found that the three-body decay and four-body radiative decay impose comparable constraints in the $\Lambda-m$ plane. However, in the mass range from $\Delta m_{\mu e}/4$ to the DM threshold, 
the muon three-body decay sets more stringent limits. The missing energy distribution of the four-body muon radiative decay is also important to determine the mass of the invisible particle.
In the case of muonium invisible decay, we found that the branching ratios from several DSEFT operators can be significantly enhanced for DM masses of tens of MeV, in contrast to the SM process involving a neutrino pair. Especially, since the SM does not predict two-body invisible decay for para-muonium, any observation of such an invisible decay would provide a compelling signature of these interactions in future low energy experiments.

\acknowledgments

We thank anonymous referees for correcting a mistake about our previous discussion concerning tau lepton decays.
We thank Yoshiki Uchida for useful discussions at the initial stage of this work. 
This work was supported by Grants No.\,NSFC-12305110 and No.\,NSFC-12035008. 

\appendix

\section{Matrix elements}
\label{app:Mi2jDMDM}
In the following, we present the complete three-body decay matrix elements for the LFV decay process, $\ell_i(p) \to \ell_j(k)+{\tt DM}(k_1) +{\tt DM}(k_2)$, in all DM scenarios. For convenience, we denote $q^\mu\equiv k_1^\mu+k_2^\mu$. Then we obtain
\begin{subequations}
\label{eq:amps}
\begin{align}
\mathcal{M}_{\ell_i \to \ell_j \phi \phi} & =\bar{u}_j\left[
\big(C_{\ell \phi}^{{\tt S},ji}
+ i \gamma_5 C_{\ell \phi}^{{\tt P},ji}\big) 
+(\slashed{k}_1-\slashed{k}_2)
\big(C_{\ell \phi}^{{\tt V},ji}+\gamma_5 C_{\ell \phi}^{{\tt A},ji}\big)\right] u_i, 
\\
\mathcal{M}_{\ell_i \to \ell_j \chi \chi} & = 
\bar{v}_\chi \big( C_{\ell \chi 1}^{{\tt S},ji} 
+ i \gamma_5 C_{\ell \chi 2}^{{\tt S},ji} \big)u_\chi (\bar{u}_j u_i)
+ \bar{v}_\chi \big( i C_{\ell \chi 1}^{{\tt P},ji} 
+\gamma_5 C_{\ell \chi 2}^{{\tt P},ji} \big) u_\chi 
(\bar{u}_j\gamma_5u_i)
\nonumber \\
&+\bar{v}_\chi \gamma_{\mu} \big(C_{\ell \chi 1}^{{\tt V},ji} 
+ \gamma_5 C_{\ell \chi 2}^{{\tt V},ji}\big)u_\chi (\bar{u}_j\gamma^{\mu}u_i) 
+ \bar{v}_\chi \gamma_{\mu} \big(C_{\ell \chi 1}^{{\tt A},ji} 
+ \gamma_5 C_{\ell \chi 2}^{{\tt A},ji} \big)u_\chi (\bar{u}_j\gamma^{\mu} \gamma_5 u_i)
\nonumber \\
&+\bar{v}_\chi \sigma_{\mu \nu} \big(C_{\ell \chi 1}^{{\tt T},ji}  
+ \gamma_5 C_{\ell \chi 2}^{{\tt T},ji}  \big)u_\chi (\bar{u}_j \sigma^{\mu \nu}u_i),
\\
\mathcal{M}^A_{\ell_i \to \ell_j X X} &= 
\epsilon^{*}_{\rho}(k_1)\epsilon^{*}_{\sigma}(k_2)\bar{u}_j\Big[ 
g^{\rho \sigma} \big(C^{{\tt S},ji}_{\ell X} + i \gamma_5 C^{{\tt P},ji}_{\ell X} \big) 
+ \sigma^{\rho \sigma} \big(iC^{{\tt T},ji}_{\ell X1}+\gamma_5 C^{{\tt T},ji}_{\ell X2}\big)
\nonumber\\ 
& +[ \gamma^{\rho}(p+k)^{\sigma} + \gamma^{\sigma}(p+k)^{\rho}] 
\big(C^{{\tt V},ji}_{\ell X1}+\gamma_5 C^{{\tt A},ji}_{\ell X1} \big)
 +i \big( \gamma^\rho k_1^{\sigma}+\gamma^\sigma k_2^{\rho} \big) 
\big(C^{{\tt V},ji}_{\ell X2}+\gamma_5 C^{{\tt A},ji}_{\ell X2}\big)
\nonumber\\
& + i\epsilon^{\mu \nu \rho \sigma }(k_2-k_1)_{\nu}\gamma_{\mu}
\big(C^{{\tt V},ji}_{\ell X3}+\gamma_5 C^{{\tt A},ji}_{\ell X3} \big)
+g^{\rho \sigma}(\slashed{k}_1-\slashed{k}_2)
\big(C^{{\tt V},ji}_{\ell X 4}+\gamma_5 C^{{\tt A},ji}_{\ell X 4} \big)
\nonumber \\
& + (\gamma^{\sigma} k_2^{\rho}- \gamma^\rho k_1^{\sigma}) 
\big(C^{{\tt V},ji}_{\ell X5}+\gamma_5 C^{{\tt A},ji}_{\ell X5} \big)
-\epsilon^{\mu \nu \rho \sigma} q_{\nu}\gamma_{\mu}
\big(C^{{\tt V},ji}_{\ell X6}+\gamma_5 C^{{\tt A},ji}_{\ell X6} \big) \Big] u_i, 
\\
\mathcal{M}^B_{\ell_i \to \ell_j X X} & = \epsilon^{*}_{\rho}(k_1)\epsilon^{*}_{\sigma}(k_2)  \bar{u}_j \Big[
2 \big( k_1^{\sigma} k_2^{\rho}- k_1\cdot k_2\, g^{\rho \sigma} \big) 
\big(\tilde{C}^{{\tt S},ji}_{\ell X 1}  + i\gamma_5 \tilde{C}^{{\tt P},ji}_{\ell X 1}  \big)
\nonumber \\
& + 2 \epsilon^{\mu \nu \rho \sigma} k_{1\mu} k_{2\nu} 
\big(\tilde{C}^{{\tt S},ji}_{\ell X 2} + i \gamma_5 \tilde{C}^{{\tt P},ji}_{\ell X 2} \big) 
\nonumber \\
&+  (k_{1\nu} g^\rho_\alpha - k_{1\alpha} g^\rho_\nu)
(k_2^\alpha g^\sigma_\mu - k_{2\mu} g^{\alpha \sigma})
\sigma^{\mu\nu}\big( i  \tilde{C}^{{\tt T},ji}_{\ell X 1}
+ \gamma_5\tilde{C}^{{\tt T},ji}_{\ell X 2}
\big) \Big] u_i.  
\end{align}
\end{subequations}
In the case of a real scalar, a Majorana fermion, or a real vector DM, one simply sets to zero the corresponding WCs associated with the operators marked with a ``$\times$'' in \cref{tab:operators}, and multiplies the above amplitudes by an additional factor of 2.

\section{Differential distributions}
\label{app:diff.rate}

We implement the amplitudes in \cref{eq:amps} into {\it Mathematica} and employ {\tt FeynCalc} package \cite{Mertig:1990an,Shtabovenko:2016sxi} to calculate the squared matrix elements, followed by the phase space integration. 
To better present the results, we define the following abbreviations, 
\begin{align}
s \equiv  (k_1+k_2)^2 =q^2, ~ ~
\kappa_f \equiv 1-{ 4m^2 \over s}, ~~
\rho_\pm\equiv(m_i\pm m_j)^2-s>0, 
\end{align}
where $m_i$, $m_j$, and $m$ represent the mass of the initial lepton, the final lepton, and the DM particle, respectively. 
Note that $\rho_\pm$ are related to the triangle function, $\lambda(x,y,x)\equiv x^2 + y^2 + z^2 - 2(xy+yz+zx)$, by
$\rho_+\rho_-=\lambda(s,m_i^2,m_j^2)$. 
The final results are summarized below. 
For the scalar DM case, 
\begin{align}
 \frac{d\Gamma_{\ell_i \to \ell_j \phi \phi}}{ds}  =&
 \frac{\sqrt{\kappa_f \rho_+ \rho_-} }{768 \pi ^3 m_i^3} \Big[ 
 3\rho_+ |C^{{\tt S},ji}_{\ell \phi}|^2
+ \kappa_f \rho_-(\rho_++3s) |C^{{\tt V},ji}_{\ell \phi}|^2\Big]
\nonumber\\
&+\Big(\rho_+\leftrightarrow \rho_-, 
C^{{\tt S},ji}_{\ell \phi}\to C^{{\tt P},ji}_{\ell \phi},
C^{{\tt V},ji}_{\ell \phi}\to C^{{\tt A},ji}_{\ell \phi} \Big).
 \label{eq:dGammadq2_S}
 \end{align}
For the fermion DM case,
\begin{align}
 \frac{d\Gamma_{\ell_i \to \ell_j \chi \chi}}{d s} & =
\frac{\sqrt{\kappa_f \rho_+\rho_-}} {384 \pi^3 m_i^3}\Big\{
3 \kappa_f s \rho_+ |C^{{\tt S},ji}_{\ell \chi 1}|^2
+ 3 s \rho_+ |C^{{\tt S},ji}_{\ell \chi 2}|^2
+(3-\kappa_f)\rho_- (\rho_+ + 3 s)
|C^{{\tt V},ji}_{\ell \chi 1}|^2
 \nonumber\\
 &+\big[ (3-\kappa_f)\rho_+\rho_- + 3(1-\kappa_f)s\rho_+
 + 6 \kappa_f s \rho_-\big]
 |C^{{\tt V},ji}_{\ell \chi 2}|^2
 \nonumber\\
 &+4\big[ 2(3-\kappa_f)\rho_+\rho_- + 3 \kappa_f s \rho_+
 + 3(3-2\kappa_f) s \rho_-\big]
 |C^{{\tt T},ji}_{\ell \chi 1}|^2 \Big\}
 \nonumber\\
 &+\Big(\rho_+\leftrightarrow \rho_-, 
 C^{{\tt S},ji}_{\ell\chi1,2} \to C^{{\tt P},ji}_{\ell\chi1,2},
 C^{{\tt V},ji}_{\ell\chi1,2} \to C^{{\tt A},ji}_{\ell\chi1,2}, 
 C^{{\tt T},ji}_{\ell \chi1} \to C^{{\tt T},ji}_{\ell\chi2}\Big)
 +\cdots,
 \label{eq:dGammadq2_F}
 \end{align}
where the dots denote the interference terms among different DSEFT interactions, which are not pertinent to our numerical analysis as we consider one single operator a time. \\
For the vector DM case A,
\begin{align}
 \frac{d\Gamma^A_{\ell_i \to \ell_j X X}}{d s} & = 
 \frac{\sqrt{\kappa_f \rho_+ \rho_- }}{768 \pi^3 (1-\kappa_f)^2 m^3_{i}s}
 \Big\{
3 (3 - 2 \kappa_f + 3 \kappa_f^2)s\rho_+ |C^{{\tt S},ji}_{\ell X}|^2 
 \nonumber\\
 & + 2\Big[ 2(3 -\kappa_f^2)\rho_+ \rho_- 
 + 3(1-\kappa_f)(3-\kappa_f) s \rho_+  
 + 6(2-\kappa_f)\kappa_f s \rho_-  
 \Big]
 |C^{{\tt T},ji}_{\ell X 1}|^2
  \nonumber\\
 & + \frac{4}{5s}\Big[ 
 (15-10\kappa_f + 3 \kappa_f^2)\rho_+^2\rho_-^2
 + 10 s\big( (1+\kappa_f)\kappa_f \rho_+^2\rho_- 
 + (3 -2\kappa_f) \rho_+ \rho_-^2 \big)
 \nonumber\\
 & 
 + 5\kappa_f s^2\big(3\kappa_f \rho_+^2 +3(1-\kappa_f)\rho_-^2
 + 4\rho_+ \rho_- \big)
 + 15\kappa_f s^3 \big( \kappa_f \rho_+ + (1-\kappa_f)\rho_- \big)
 \Big]
 |C^{{\tt V},ji}_{\ell X 1}|^2
 \nonumber\\
 & + 4 \kappa_f s \Big[ 
(1 + 2 \kappa_f) \rho_+ \rho_- + 3 \kappa_f s \rho_+ + 3 (1 - \kappa_f) s \rho_- \Big]
 |C^{{\tt V},ji}_{\ell X 2}|^2
 \nonumber\\
 &+ 2 (1-\kappa_f) \kappa_f s \Big[ 
 (3-\kappa_f) \rho_+ \rho_- + 3 (1-\kappa_f) s \rho_+ + 6 \kappa_f s \rho_- \Big]
 |C^{{\tt V},ji}_{\ell X 3}|^2
 \nonumber\\
 &+  (3-2\kappa_f + 3\kappa_f^2 )\kappa_f s
\rho_- (\rho_+ + 3 s)  |C^{{\tt V},ji}_{\ell X 4}|^2
+ 4 (2 -\kappa_f)\kappa_f s   
\rho_- (\rho_+ + 3 s)  |C^{{\tt V},ji}_{\ell X 5}|^2
 \nonumber\\
 &
+ 2 (1 -\kappa_f)(3-\kappa_f)s   
\rho_- (\rho_+ + 3 s)  |C^{{\tt V},ji}_{\ell X 6}|^2
 \Big\}
  \nonumber\\
 &+\Big(\rho_+\leftrightarrow \rho_-, 
 C^{{\tt S},ji}_{\ell X} \to C^{{\tt P},ji}_{\ell X},
 C^{{\tt T},ji}_{\ell X1} \to C^{{\tt T},ji}_{\ell X2}, 
 C^{{\tt V},ji}_{\ell X1-6} \to C^{{\tt A},ji}_{\ell X1-6}\Big) 
  +\cdots.
 \label{eq:dGammadq2_VA}
 \end{align}
For the vector DM case B,
\begin{align}
 \frac{d\Gamma^B_{\ell_i \to \ell_j X X}}{d s}& = 
 \frac{s\sqrt{\kappa_f \rho_+\rho_-}}{6144\pi^3 m^3_i } 
 \Big\{
6 (3 +2 \kappa_f + 3\kappa_f^2) s \rho_+  
|\tilde{C}^{{\tt S},ji}_{\ell X 1}|^2
+ 48 \kappa_f s \rho_+ 
|\tilde{C}^{{\tt S},ji}_{\ell X 2}|^2
 \nonumber
 \\
 &+\Big[ 2(3+ 6\kappa_f -\kappa_f^2)\rho_+\rho_- 
 + 3(3+\kappa_f^2)s \rho_+ + 6(3-\kappa_f)\kappa_f s \rho_- \Big] 
 |\tilde{C}^{{\tt T},ji}_{\ell X 1}|^2\Big\}
\nonumber\\
&+\Big(\rho_+\leftrightarrow \rho_-, 
\tilde C^{{\tt S},ji}_{\ell X1,2} \to \tilde C^{{\tt P},ji}_{\ell X1,2}, 
\tilde C^{{\tt T},ji}_{\ell X1} \to  \tilde C^{{\tt T},ji}_{\ell X2}\Big)
 +\cdots.
 \label{eq:dGammadq2_VB}
 \end{align}
In the case of identical final-state DM particles, an additional factor of 2 should be included. This is a joint result from a factor 2 in the amplitude and a factor $1/2$ from phase space.

\bibliography{refs.bib}{}
\bibliographystyle{JHEP}
\end{document}